\RequirePackage[dvipsnames,hyperref,table]{xcolor}
\documentclass[9pt,twocolumn,twoside]{osajnl}

\journal{josaa} 

\setboolean{shortarticle}{true}

\ifthenelse{\boolean{shortarticle}}{\colorlet{color2}{color2b}}{\colorlet{color2}{color2}} 
\renewcommand*{\journalshorttype}{Tutorial} 

\usepackage[utf8]{inputenc}
\usepackage[normalem]{ulem}

\newcommand{\Hide}[1]{}

\usepackage{dsfont}

\usepackage{graphicx}
\graphicspath{{figs/}}

\usepackage{amsmath,amsfonts,amssymb}
\usepackage{mathtools}
\usepackage[squaren,Gray]{SIunits}
\usepackage{xspace}

\usepackage{array}
\usepackage{booktabs}
\colorlet{tableheadcolor}{gray!25}
\colorlet{tablerowcolor}{gray!12.5}
\newcommand{\TopRule}{\specialrule{\heavyrulewidth}{0pt}{0pt}}
\newcommand{\BottomRule}{\specialrule{\lightrulewidth}{0pt}{0pt}}
\newcolumntype{P}[1]{>{\raggedright}p{#1}}
\newcolumntype{M}[1]{>{\raggedright}m{#1}}

\newcommand*{\etal}{\emph{et al.}\xspace}
\newcommand*{\etc}{\emph{etc.}\xspace}
\newcommand*{\eg}{\emph{e.g.}\xspace}
\newcommand*{\ie}{\emph{i.e.}\xspace}

\DeclareSymbolFont{bbold}{U}{bbold}{m}{n}
\DeclareSymbolFontAlphabet{\mathbbold}{bbold}
\RequirePackage{textcomp}  

\DeclarePairedDelimiterX{\Paren}[1]{(}{)}{#1}
\DeclarePairedDelimiterX{\Brace}[1]{\{}{\}}{#1}
\DeclarePairedDelimiterX{\Brack}[1]{[}{]}{#1}
\DeclarePairedDelimiterX{\Abs}[1]{\rvert}{\lvert}{#1}
\DeclarePairedDelimiterX{\Norm}[1]{\lVert}{\rVert}{#1}
\DeclarePairedDelimiterX{\Avg}[1]{\langle}{\rangle}{#1}
\DeclarePairedDelimiterX{\Round}[1]{\lfloor}{\rceil}{#1}
\DeclarePairedDelimiterX{\Floor}[1]{\lfloor}{\rfloor}{#1}
\DeclarePairedDelimiterX{\Ceil}[1]{\lceil}{\rceil}{#1}
\DeclarePairedDelimiterX{\Inner}[2]{\langle}{\rangle}{#1,#2}

\DeclarePairedDelimiterXPP{\List}[3]{}{\{}{\}}{_{#2,\ldots,#3}}{#1}

\newcommand*{\delimsize}{}
\newcommand*{\given}{\,\vert\,} 
\newcommand*{\Given}{\:\delimsize\vert\:} 

\newcommand*{\mathd}{\mathrm{d}}
\newcommand*{\mathe}{\mathrm{e}}
\newcommand*{\mathi}{\mathrm{i}}

\renewcommand*{\Re}{\operatorname{Re}}
\renewcommand*{\Im}{\operatorname{Im}}
\DeclareMathOperator*{\argmin}{arg\,min}
\DeclareMathOperator*{\argmax}{arg\,max}
\DeclareMathOperator{\arc}{arc}

\DeclareMathOperator{\Diag}{diag}

\DeclareMathOperator{\Sign}{sign}

\DeclareMathOperator{\Var}{Var}
\DeclareMathOperator{\Cov}{Cov}
\DeclareMathOperator{\Expect}{E}

\newcommand*{\Set}[1]{\mathbb{#1}}
\newcommand*{\Reals}{\Set{R}}

\newcommand*{\Complexes}{\Set{C}}

\newcommand*{\Tag}[1]{\mathsf{#1}}

\newcommand*{\estim}[1]{\widehat{#1}}
\newcommand*{\FT}[1]{\widetilde{#1}}

\newcommand*{\V}[1]{\boldsymbol{#1}}   
\newcommand*{\M}[1]{\mathbf{#1}}       

\newcommand*{\TransposeLetter}{\mathrm{t}}
\newcommand*{\T}{^{\TransposeLetter}}

\newcommand*{\MDot}{\mathord{\,\mathchar"2201\,}}

\newcommand*{\QuadTerm}[2]{#2\T\MDot#1\MDot#2}
\newcommand{\ArrayWithDelimiters}[4]{%
  \left #1\begin{array}{#2}#4\end{array}\right #3}
\newcommand{\Matrix}[2]{\ArrayWithDelimiters{(}{#1}{)}{#2}}

  \def\clap#1{\hbox to 0pt{\hss#1\hss}}
  
  \def\mathrlap{\mathpalette\mathrlapinternal}

  \def\mathrlapinternal#1#2{\rlap{$\mathsurround=0pt#1{#2}$}}

\makeatletter

\newcommand*{\@SymbolBuilderWithLabel}[4]{%
  \begingroup \escapechar\m@ne\xdef\my@tempa{\string#1}\endgroup
  \expandafter\@ifundefined{\my@tempa}{%
    \def\my@tempb{[} 
    \expandafter\edef\csname\my@tempa\endcsname{%
      \noexpand\@ifnextchar\my@tempb%
          {\csname\my@tempa @b\endcsname}%
          {\csname\my@tempa @a\endcsname}%
    }
    \expandafter\def\csname\my@tempa @a\endcsname{%
      {#2}_{#4}%
    }
    \expandafter\def\csname\my@tempa @b\endcsname[##1]{%
      {{#3}^{#4}_{##1}}%
    }
  }{\@latex@error{\noexpand#1is already defined}\@ehc}%
}
\newcommand*{\Variable}[3]{
  \begingroup \escapechar\m@ne\xdef\my@tempa{\string#1}\endgroup
  \expandafter\@ifundefined{\my@tempa}{%
    \def\my@tempb{[}
    \expandafter\edef\csname\my@tempa\endcsname{%
      \noexpand\@ifnextchar\my@tempb%
          {\csname\my@tempa @b\endcsname}%
          {\csname\my@tempa @a\endcsname}%
    }
    \expandafter\def\csname\my@tempa @a\endcsname{%
      #2%
    }
    \expandafter\def\csname\my@tempa @b\endcsname[##1]{%
      {{#3}_{##1}}%
    }
  }{\@latex@error{\noexpand#1is already defined}\@ehc}%
}

\newcommand*{\SymbolWithLabel}[3]{%
  \@SymbolBuilderWithLabel{#1}{#2}{#2}{#3}}

\newcommand*{\VectorSymbolWithLabel}[3]{%
  \@SymbolBuilderWithLabel{#1}{\V{#2}}{#2}{#3}}
\newcommand*{\VectorSymbol}[2]{%
  \Variable{#1}{\V{#2}}{#2}}

\newcommand*{\MatrixSymbolWithLabel}[3]{%
  \@SymbolBuilderWithLabel{#1}{\M{#2}}{#2}{#3}}
\newcommand*{\MatrixSymbol}[2]{%
  \Variable{#1}{\M{#2}}{#2}}

\makeatother

\newcommand*{\Fdata}{f_\Tag{data}}
\newcommand*{\Fprior}{f_\Tag{prior}}
\newcommand*{\PDF}{\Pr}

\VectorSymbol{\Param}{x}
\Variable{\BestParam}{\estim{\V x}}{\estim{x}}
\Variable{\DefSolution}{\V x^\Tag{def}}{x^\Tag{def}}
\Variable{\Hyper}{\V\omega}{\omega}
\VectorSymbol{\Data}{y}
\VectorSymbol{\Noise}{\varepsilon}
\MatrixSymbol{\ModelOp}{H} 
\newcommand*{\ModelFn}{\mathcal{H}} 

\newcommand*{\atmphase}{\phi^\Tag{atm}} 
\newcommand*{\instphase}{\psi^\Tag{inst}} 
\newcommand*{\objphase}{\varphi} 
\newcommand*{\closurephase}{\beta} 
\newcommand*{\totalflux}{\varrho} 
\newcommand*{\diffvis}{D} 

\newcommand*{\vocab}[1]{\textbf{#1}\xspace}
\newcommand*{\noun}[1]{\textsc{#1}\xspace}
\newcommand*{\BSMEM}{BSMEM\xspace}
\newcommand*{\MIRA}{MiRA\xspace}
\newcommand*{\CLEAN}{\noun{Clean}}
\newcommand*{\WISARD}{\noun{Wisard}}
\newcommand*{\PAINTER}{\noun{Painter}}
\newcommand*{\SQUEEZE}{\noun{Squeeze}}
\newcommand*{\SPARCO}{\noun{Sparco}}
\newcommand*{\MACIM}{\noun{Macim}}
\newcommand*{\IRBIS}{\noun{IRBis}}
\newcommand*{\WIPE}{\noun{Wipe}}

\newcommand*{\BispectrumTag}{\Tag{bisp}}
\newcommand*{\PowerspectrumTag}{\Tag{pow}}
\newcommand*{\ClosurePhaseTag}{\Tag{clos}}
\newcommand*{\VisibilityTag}{\Tag{vis}}

\newcommand*{\Telescopes}{\mathcal{T}}

\title{Principles of image reconstruction in optical interferometry: tutorial}

\author[1]{Éric Thiébaut}
\author[2]{John Young}

\affil[1]{Univ.\ Lyon, Univ.\ Lyon 1, ENS de Lyon, CNRS, Centre de Recherche
  Astrophysique de Lyon UMR5574, F-69230, Saint-Genis-Laval, France}

\affil[2]{Univ. Cambridge, Cavendish Laboratory, JJ Thomson Avenue, Cambridge CB3 0HE, UK}

\affil[*]{Corresponding author: eric.thiebaut@univ-lyon1.fr}

\dates{Received 2 March 2017; revised 6 April 2017; accepted 6 April 2017; posted 19 April 2017 (Doc. ID 287964); published 15 May 2017}

\ociscodes{100.3020; 
           100.3190; 
           100.3175; 
          }

\doi{\url{http://dx.doi.org/10.1364/JOSAA.34.000904}}

\begin{abstract}
This paper provides a general introduction to the problem of image
reconstruction from interferometric data.  A simple model of the
interferometric observables is given and the issues arising from sparse
Fourier data are discussed. The effects of various regularizations are
described. In the proposed general framework, most existing algorithms can
be understood.  For an astronomer, such an understanding is crucial not
only for selecting and using an algorithm but also to ensure correct
interpretation of the resulting image.
\end{abstract}

\setboolean{displaycopyright}{true}

\begin{document}

\maketitle
\thispagestyle{fancy}

\ifthenelse{\boolean{shortarticle}}{\ifthenelse{\boolean{singlecolumn}}{\abscontentformatted}{\abscontent}}{}

\section{Introduction}

Astronomical interferometers sample the so-called \emph{visibility} which
is the Fourier transform of the angular brightness distribution of the
observed object \cite{Monnier-2003-interferometry,
Quirrenbach-2009-interferometry}.  These instruments achieve unrivaled
angular resolution but only provide an irregular and sparse coverage of the
Fourier frequencies. Forming an image from the interferometric observables
is then not a simple matter of performing an inverse Fourier transform of
the data. Additional \emph{a priori} assumptions about the regularity (\ie
``smoothness'') or the simplicity of the object are needed in order to
interpolate the voids in the frequency coverage of the measurements. Image
restoration is then an inverse problem whose solution is a compromise
between fitting the available data and keeping the image as regular (or
simple) as possible
\cite{Tarantola_Valette-1982-inverse_problems_quest_for_information,
Titterington-1985-regularization, Tarantola-1987-inverse_problem_theory,
Demoment-1989-image_reconstruction,
Thiebaut_Giovannelli-2010-interferometry}. In radio-astronomy, image
reconstruction algorithms have a long and successful history
\cite{Hogbom-1974-CLEAN, Narayan_Nityananda-1986-maximum_entropy_review,
Thompson_et_al-2017-radio_astronomy} with renewed interest due to the
emergence of the \emph{compressive sensing} theory
\cite{Candes_Wakin-2008-compressive_sampling}. There are additional issues
at optical wavelengths which make the image reconstruction much more
difficult than in radio-astronomy. First, the smaller number of recombined
telescopes yields a much sparser frequency coverage. Second, to cancel the
effects of the turbulence, non-linear observables such as the power
spectrum and the closure phase must be formed from the visibilities. With
such observables more information is lost and recovering an image becomes a
more difficult non-convex inverse problem.

Although there now exist a number of good image reconstruction algorithms
dedicated to the processing of optical interferometry data
\cite{Baron-2016-image_reconstruction_overview}, they are not unsupervised
black boxes magically producing an image without user control.  To use
these algorithms successfully, astronomers should be accustomed to the
underlying general principles of image reconstruction from sparse data.
This is of prime importance for choosing the prior constraints and setting
the tuning parameters of the methods.  This knowledge is also helpful for
choosing between the available algorithms (most being publicly available)
and their numerous options. It is also necessary to understand how
interferometric observables are measured and their intrinsic limitations,
some of which directly impact the image reconstruction and the uniqueness
of the solution. It must be stressed that assuming the existence of a
single best image for a given data set is, owing to the small number of
measurements, a naive belief.  As a consequence there is not a single best
algorithm and the observer should try different methods and settings to
analyze her/his data. This, again, requires some understanding of the
underlying recipes. Although we encourage the reader to dive into the vast
literature of inverse problems and image reconstruction, this article
aspires to provide a didactic and yet detailed introduction to these
methods. The general framework of inverse problems has already been used to
formally describe many algorithms for image reconstruction from
interferometric data \cite{Thiebaut-2009-interferometric_imaging,
Thiebaut_Giovannelli-2010-interferometry}.  A similar framework is used in
this paper but with an emphasis on the specific issues in optical
interferometry and with an overview of the most recent algorithms in this
context.  The benefits of the most common regularization methods when
applied to very sparse interferometric data have already been studied
\cite{Renard_et_al-2011-regularization}.  Here we present an updated list
of regularization methods (\eg the ones implementing sparsity priors) under
a more general formulation which is consistent with the adopted framework.

This paper is organized as follows.  Starting with a very simple
description of an interferometer, we derive the kind of measurements that
are provided by such an instrument.  We then introduce the principles of
image restoration from incomplete data.  We explain how to quantitatively
compare the image and the data.  We detail the available regularizations
and their properties. Finally we provide a short review of the algorithms
that have been developed for optical interferometry.

\section{A simple interferometer}

\begin{figure}
  \centering
  \includegraphics[width=80mm]{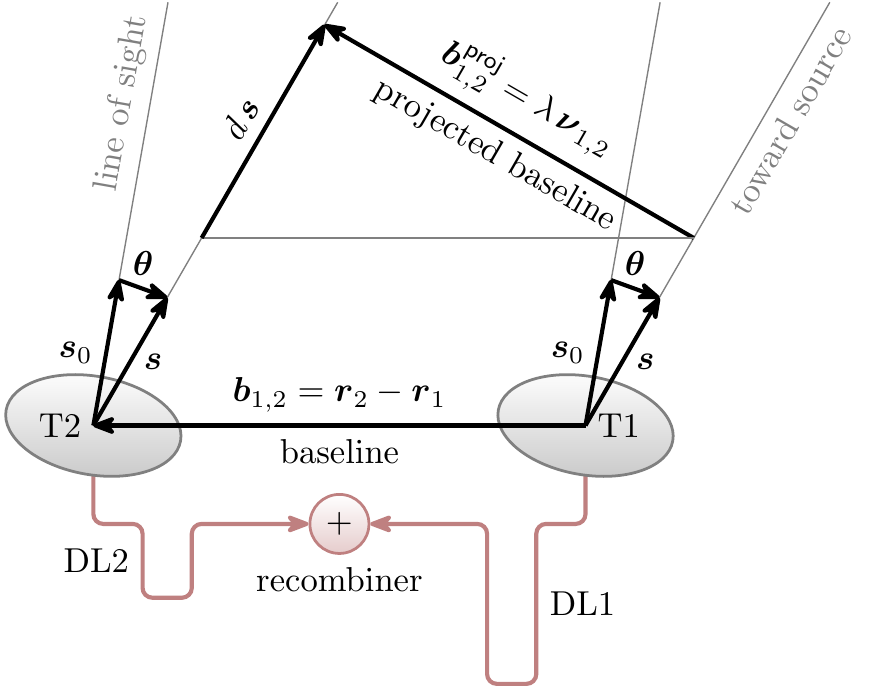}
  \caption{A simple 2-telescope stellar interferometer.  The two telescopes
  are denoted by T1 and T2 and the corresponding delay lines are denoted by
  DL1 and DL1.}\label{fig:interferometer}
\end{figure}

A stellar interferometer consists of two or more telescopes which sample
the light wavefronts from a celestial source at spatially-separated
locations. These wavefront samples are interfered in some place, the
\emph{recombiner}, where a detector is located.  Delay lines are inserted
in the optical path to compensate for geometrical optical delays and to
introduce small phase shifts. Figure~\ref{fig:interferometer} shows the
layout of a typical 2-telescope interferometer.  For the sake of
simplicity, we consider in the following the case of a mono-mode stellar
interferometer observing an object of small angular size.

\subsection{Observed Signal}

What can be  measured by an interferometer is determined by the intensity
of the light at the recombiner for a given observed object. The object is
characterized by $A_\Tag{obj}(\V\theta)$ the complex amplitude of the light
collected by the telescopes of the interferometer in the direction
$\V\theta = \V s - \V s_0$ relative to the center of the field of view ($\V
s$ and $\V s_0$ being unit vectors pointing in the considered direction and
to the center of the field of view respectively, see
Fig.~\ref{fig:interferometer}).  For a mono-mode stellar interferometer
\cite{Froehly-1981-interferometry_through_optical_fibers,
Shaklan_Roddier-1987-single_mode,
du_Foresto_et_al-1997-fiber_stellar_interferometer}, the monochromatic
complex amplitude transmitted by a given telescope is a scalar quantity.
The general expression of the complex amplitude transmitted by the $j$-th
telescope to the recombiner is therefore given by:
\begin{equation}
  \label{eq:ampl}
  A_{j} = \iint T_{j}(\V\theta) \, A_\Tag{obj}(\V\theta)
  \, \mathd^2\V\theta \, ,
\end{equation}
with $T_{j}(\V\theta)$ a complex amplitude transmission. The total complex
amplitude at the recombiner is just the sum of the contributions of the
recombined telescopes:
\begin{equation}
  A_\Telescopes = \sum_{\mathrlap{j\in\Telescopes}} A_{j}
  = \iint A_\Tag{obj}(\V\theta) \, \sum_{\mathrlap{j\in\Telescopes}}
   T_{j}(\V\theta) \, \mathd^2\V\theta \, ,
\end{equation}
where $\Telescopes$ is the list of recombined telescopes.  At optical
wavelengths, only the squared modulus of the complex amplitude averaged
over a few wave periods can be measured. This quantity is known as the
intensity which, for a detector placed at the recombiner, is given by:
\begin{align*}
  I_\Telescopes
  &= \Avg[\big]{\Abs{A_\Telescopes}^2} \notag\\
  &= \iiiint \Avg[\big]{A_\Tag{obj}(\V\theta)\,A_\Tag{obj}^\star(\V\theta')}
  \sum_{\mathrlap{(j,j')\in\Telescopes^2}}
  T_{j}(\V\theta) \, T_{j'}^{\star}(\V\theta')
  \, \mathd^2\V\theta \, \mathd^2\V\theta' \, ,
\end{align*}
where $\Avg{\ }$ denotes the considered time averaging and the exponent
$\star$ denotes complex conjugation. For an incoherent source object,
averaging the product of complex amplitudes yields:
\begin{align}
  \label{eq:incoherent-cross-product}
  \Avg[\big]{A_\Tag{obj}(\V\theta)\,A_\Tag{obj}^\star(\V\theta')} &=
  \begin{cases}
  \Avg[\big]{A_\Tag{obj}(\V\theta)}\,\Avg[\big]{A_\Tag{obj}^\star(\V\theta')} = 0&
  \text{if $\V\theta \not= \V\theta'$} \\
  \Avg[\big]{\Abs{A_\Tag{obj}(\V\theta)}^ 2} = I_\Tag{obj}(\V\theta)&
    \text{if $\V\theta = \V\theta'$} \\
  \end{cases} \notag \\
  &= I_\Tag{obj}(\V\theta) \, \delta(\V\theta' - \V\theta) \, ,
\end{align}
where $I_\Tag{obj}(\V\theta)$ is the brightness distribution of the object.
Then:
\begin{equation}
  I_\Telescopes
  = \iint I_\Tag{obj}(\V\theta)
  \sum_{\mathrlap{(j,j')\in\Telescopes^2}}
  T_{j}(\V\theta) \, T_{j'}^{\star}(\V\theta) \, \mathd^2\V\theta \, .
  \label{eq:Irec-1}
\end{equation}
In order to further simplify this equation, we shall consider an object of
small angular size compared to the \emph{primary beam}
\cite{Thompson_et_al-2017-radio_astronomy} of the light collectors (here
the telescopes coupled with their mono-mode filters).  Then, the modulus of
the amplitude transmission $T_{j}(\V\theta)$ does not depend on $\V\theta$:
\begin{equation}
  \label{eq:ampl-trans}
  T_j(\V\theta) = 
  \tau_j\,
  \mathe^{\mathi\,\phi_{j}(\V\theta)} \, ,
\end{equation}
with $\tau_j = \Abs{T_j(\V\theta)} \ge 0$ and $\phi_{j}(\V\theta)$ a phase
term. For such a small object, the intensity at the recombiner becomes:
\begin{equation}
  I_\Telescopes
  = \iint I_\Tag{obj}(\V\theta)
  \Brack[\Big]{
    \sum_{j \in \Telescopes} \tau_j^2 +
    2 \, \sum_{j < j'} \tau_j \, \tau_{j'} \,
    \cos\Paren[\big]{\Delta\phi_{j,j'}(\V\theta)}
  } \, \mathd^2\V\theta \, .
  \label{eq:Irec-2}
\end{equation}
with $\Delta\phi_{j,j'}(\V\theta) = \phi_{j'}(\V\theta) -
\phi_{j}(\V\theta)$.

The phase difference $\Delta\phi_{j,j'}(\V\theta)$ has three different
contributions: (i) a quasi-static instrumental part due to the optics; (ii)
a geometrical optical path delay (denoted by $d$ on
Fig.~\ref{fig:interferometer}); (iii) an unknown random phase due to the
turbulence.  For a mono-mode stellar interferometer, the instrumental phase
does not depend on the direction of the incident wave.  Provided the
angular size of the object is smaller than the isoplanatic patch
\cite{Roddier_et_al-1982-isoplanatic_patch}, the turbulent phase term is
also independent of the direction of the incident wave.  In that case, only
the geometrical optical path delay depends on $\V\theta$ (or $\V s$).  More
specifically, the geometrical optical path delay of telescope $j'$ relative
to telescope $j$ is:
\begin{equation}
  d = \Inner{\V s}{\V b_{j,j'}}
\end{equation}
with $\V b_{j,j'} = \V r_{j'} - \V r_{j}$ the \emph{baseline} which
is the difference between the respective positions $\V r_{j}$ and $\V
r_{j'}$ of the 2 telescopes.  Here $\Inner{\cdot}{\cdot}$ denotes the
scalar product between two vectors.  Introducing
$\Delta\phi_{j,j'}(\V0)$ and $d_0 = \Inner{\V s_0}{\V b_{j,j'}}$ the
phase difference and the geometrical delay for the center of the field of
view and since $\V s = \V s_0 + \V\theta$, the phase difference between the
two telescopes is:
\begin{align}
  \Delta\phi_{j,j'}(\V\theta)
  &= \Delta\phi_{j,j'}(\V0) - \frac{2\,\pi}{\lambda}\,(d - d_0) \notag \\
  &= \Delta\phi_{j,j'}(\V0)
  - \frac{2\,\pi}{\lambda}\,\Inner{\V \theta}{\V b_{j,j'}} \, ,
\end{align}
where $\lambda$ is the wavelength. For a small field of view, $\V\theta$ is
in the plane which is tangential to the celestial sphere at the center of
the field of view.  Let $\V b^\Tag{proj}_{j,j'}$ be the projection of
the baseline $\V b_{j,j'}$ on this plane and define:
\begin{equation}
  \label{eq:spatial-frequency}
  \V\nu_{j,j'} = \V b^\Tag{proj}_{j,j'}/\lambda \, ,
\end{equation}
then:
\begin{equation}
  \Inner{\V\theta}{\V b_{j,j'}/\lambda}
  = \Inner{\V\theta}{\V b^\Tag{proj}_{j,j'}/\lambda}
  = \Inner{\V\theta}{\V\nu_{j,j'}} \, .
\end{equation}
The phase difference $\Delta\phi_{j,j'}(\V0)$ for the center of the
field of view accounts for an instrumental phase $\instphase_{j,j'}$ and
for random phase terms $\atmphase_{j}$ due to the atmospheric turbulence
above each telescope:
\begin{equation}
  \Delta\phi_{j,j'}(\V0) = \instphase_{j,j'} + \atmphase_{j'} - \atmphase_{j} \, .
\end{equation}
The instrumental phase $\instphase_{j,j'}$ must be calibrated and will be
assumed known.  The random atmospheric phases $\atmphase_{j}$ and $\atmphase_{j'}$ can also be calibrated --
despite the fact that they vary much faster -- if a phase reference such as
a nearby star is available \cite{Delplancke_at_al-2003-Prima},  but most of
the time this is not the case for current optical interferometers.
Combining the above relations yields the phase difference in the direction
$\V\theta$:
\begin{equation}
  \label{eq:phase-shift}
  \Delta\phi_{j,j'}(\V\theta)
  = \instphase_{j,j'} + \atmphase_{j'} - \atmphase_{j}
  - 2\,\pi\,\Inner{\V\theta}{\V\nu_{j,j'}} \, .
\end{equation}

For a 2-telescope interferometer, say $\Telescopes = (j_1,j_2)$,
substituting this expression into Eq.~(\ref{eq:Irec-2}) and simplifying
yields:
\begin{align}
  I_{j_1,j_2}
  &= \Paren[\big]{\tau_{j_1}^2 + \tau_{j_2}^2}\,\FT{I}_\Tag{obj}(\V0)
  \notag \\
  &\quad + 2\,\tau_{j_1}\,\tau_{j_2}\,\Re\Paren[\Big]{
     \FT{I}_\Tag{obj}(\V\nu_{j_1,j_2})\,\mathe^{\mathi\,\Paren{
       \instphase_{j_1,j_2} + \atmphase_{j_2} - \atmphase_{j_1}
     }
   }
  } \, ,
  \label{eq:fringe-pattern}
\end{align}
where $\FT{I}_\Tag{obj}(\V\nu)$ is the Fourier transform of the brightness
distribution of the object at the spatial frequency $\V\nu$:
\begin{equation}
  \label{eq:Fourier-transform}
  \FT{I}_\Tag{obj}(\V\nu) = \iint I_\Tag{obj}(\V\theta) \,
  \mathe^{-\mathi\,2\,\pi\,\Inner{\V\theta}{\V\nu}} \, \mathd^2\V\theta \, .
\end{equation}
Equation~(\ref{eq:fringe-pattern}) shows that, for given telescope
positions relative to the observed object and during times short enough to
freeze the turbulence effects, $I_{j_1,j_2}$ as a function of the
instrumental phase $\instphase_{j_1,j_2}$ displays a fringe pattern much
like that in Young's double slit experiment.  A quantity of interest which
can be extracted from this fringe pattern is the so-called \vocab{coherent
flux} \cite{Tatulli_et_al-2007-interferometric_data_reduction,
Gordon_Buscher-2012-interferometric_noise}:
\begin{align}
  C_{j_1,j_2}
  &= \tau_{j_1} \, \tau_{j_2} \,
     \FT{I}_\Tag{obj}(\V\nu_{j_1,j_2})\,
     \mathe^{\mathi\,\Paren{\atmphase_{j_2} - \atmphase_{j_1}}}
  \, .
  \label{eq:coherent-flux}
\end{align}
This shows that a 2-telescope interferometer gives access to the Fourier
transform of the brightness distribution at a spatial frequency equal to
the projected baseline divided by the wavelength, see
Eq.~(\ref{eq:spatial-frequency}).  This ability however depends on whether
the complex gains $\tau_j\,\exp\Paren{\mathi\,\atmphase_j}$ are known as
discussed next.


\subsection{Coverage of the frequency plane and index notation}

\begin{figure*}[t]
  \centering
  \includegraphics[height=110mm]{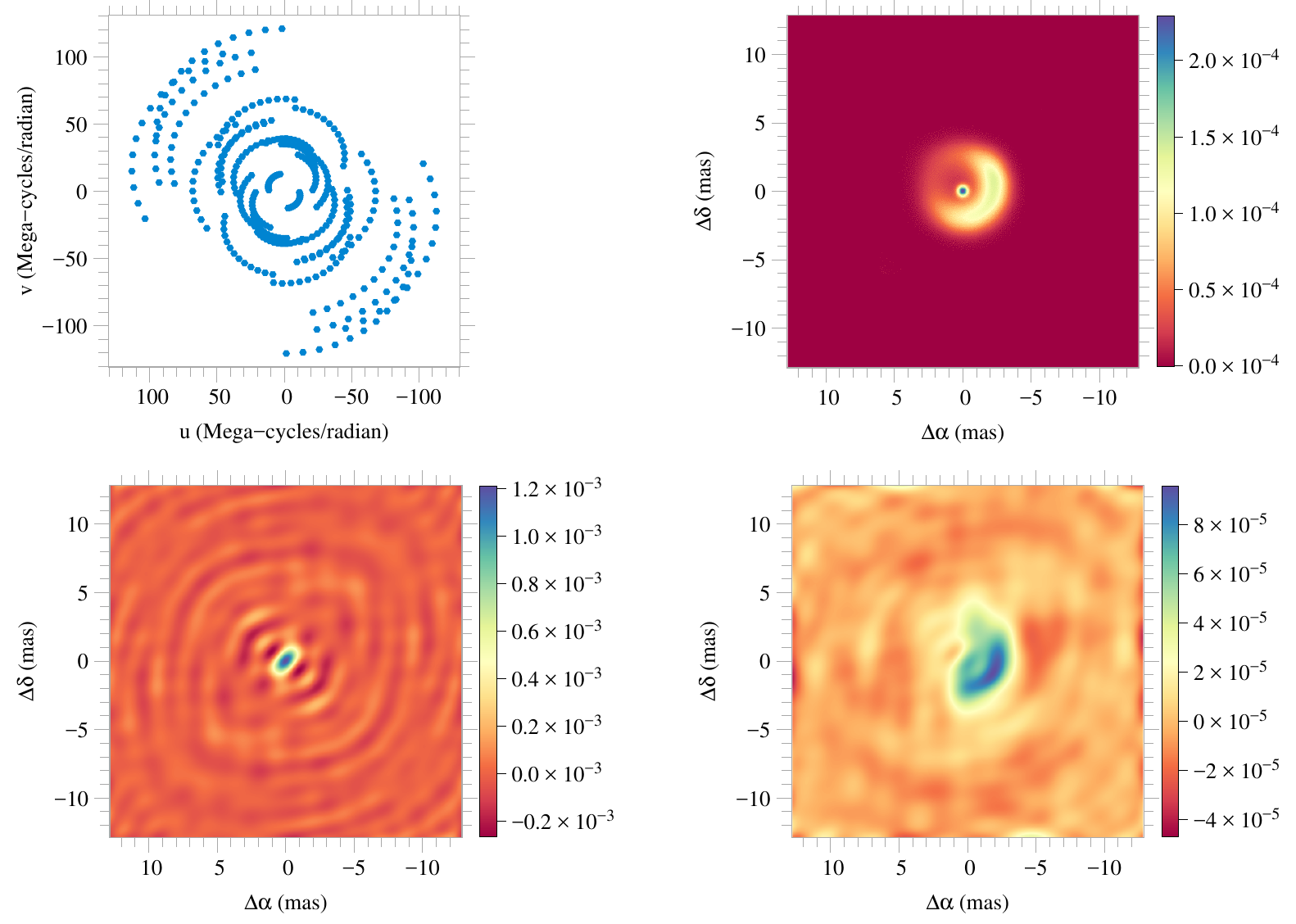}
  \caption{Top left: simulated frequency coverage sampled with a
  six-station Navy Prototype Optical Interferometer (NPOI). Top right:
  model of LkHa-101 representing the observed object. Bottom left: dirty
  beam.  Bottom right: dirty image. Object model and frequency coverage are
  from the \emph{2004 Interferometric Beauty Contest}
  \protect\cite{Lawson_et_al-2004-image_beauty_contest}.}
  \label{fig:dirty-map}
\end{figure*}

For image reconstruction, measuring a single spatial frequency is obviously
insufficient. The coverage of the frequency plane can be improved by
collecting measurements obtained with more baselines (with more than two
telescopes), at different wavelengths and at different times (by moving the
telescopes or just because the projected baselines change due to the Earth
rotation). In practice however, this sampling remains very sparse (see
top-left panel of Fig.~\ref{fig:dirty-map}) and not all frequencies are
measured.  This is one of the key issues for successful image
reconstruction in this context.


The notation in Eq.~(\ref{eq:spatial-frequency}), which only indicates the
two interfering telescopes, is insufficient to account for the temporal
variation of the projected baselines and for spectral dependence of the
measurements.  We should denote as $\V\nu_{j_1,j_2,\ell,m}$ the spatial
frequency sampled by the pair of telescopes $(j_1,j_2)$ in a spectral
channel at effective wavelength $\lambda_\ell$ and at exposure time $t_m$.
As interferences must be simultaneous, measured quantities should also be
indexed to reflect that.  For instance $C_{j_1,j_2,\ell,m}$ denotes the
coherent flux of the interferences between the telescopes $j_1$ and $j_2$
in the $\ell$-th spectral channel and during the $m$-th exposure.  When
this distinction is not necessary and to simplify the notation, we will
simply denote a spatial frequency sampled by the data as $\V\nu_k$ and the
coherent flux at that frequency as $C_{k}$.  In the following, we maintain
a consistent index notation ($j$ for a telescope, $k$ for a datum or a
frequency, $\ell$ for a spectral channel, $m$ for an exposure, $n$ for a
pixel, \emph{etc.})  so that there should be no ambiguities.

\subsection{Interferometric observables}

When the effects of the turbulence are stable or measured in real-time, the
terms $\tau_j\,\exp\Paren{\mathi\,\atmphase_j}$ can be calibrated (or
compensated) to directly infer $\FT{I}_\Tag{obj}(\V\nu_{j_1,j_2})$ from
the measured coherent flux. Otherwise, photometric channels may be used to
estimate the flux transmitted by a given telescope:
\begin{equation}
  \label{eq:photometric-flux}
  P_j = \tau_{j}^2 \,\FT{I}_\Tag{obj}(\V0) \, ,
\end{equation}
and hence, in order to remove the varying amplitude transmissions $\tau_j$,
measure the quantity:
\begin{equation}
  \label{eq:measured-visibility}
  V_{j_1,j_2} = \frac{C_{j_1,j_2}}{\sqrt{P_{j_1}\,P_{j_2}}}
  = \FT{I}(\V\nu_{j_1,j_2})\,
  \mathe^{\mathi\,\Paren{\atmphase_{j_2} - \atmphase_{j_1}}}
  \, ,
\end{equation}
where $\FT{I}(\V\nu) = \FT{I}_\Tag{obj}(\V\nu)/\FT{I}_\Tag{obj}(\V0)$ is
the Fourier transform of the normalized brightness distribution of the
object:
\begin{equation}
  \label{eq:normalized-brightness-distribution}
  I(\V\theta)
  = \frac{I_\Tag{obj}(\V\theta)}{
    \displaystyle
    \iint I_\Tag{obj}(\V\theta')\,\mathd^2\V\theta'
  } \, .
\end{equation}

The observable $V_{j_1,j_2}$ in Eq.~(\ref{eq:measured-visibility}) still
depends on an unknown phase shift $\atmphase_{j_2} - \atmphase_{j_1}$ due
to the turbulence.  In order to get rid of this nuisance phase shift, the
\vocab{power spectrum} can be measured:
\begin{equation}
  \label{eq:powerspectrum}
  S_{j_1,j_2} = \Abs[\big]{V_{j_1,j_2}}^2
  = \Abs[\big]{\FT{I}(\V\nu_{j_1,j_2})}^2 \, .
\end{equation}
The power spectrum only carries information about the modulus of the
Fourier transform of the brightness distribution.  Given the simultaneous
coherent fluxes for the 3 baselines formed by 3 different telescopes
($j_1$, $j_2$ and $j_3$), the \vocab{bispectrum}
\cite{Lohmann_et_al-1983-triple_correlation} can be formed:
\begin{equation}
  \label{eq:bispectrum}
  B_{j_1,j_2,j_3} = V_{j_1,j_2}\,V_{j_2,j_3}\,V_{j_3,j_1}
  = \FT{I}(\V\nu_{j_1,j_2}) \,
    \FT{I}(\V\nu_{j_2,j_3}) \,
    \FT{I}(\V\nu_{j_3,j_1}) \, ,
\end{equation}
to get information about the phase of the Fourier transform of the
brightness distribution.  Some image reconstruction algorithms
consider the \vocab{closure phase} which is the phase of the bispectrum:
\begin{align}
  \closurephase_{j_1,j_2,j_3}
  &= \arg\Paren{ B_{j_1,j_2,j_3}} \notag \\
  &= \objphase(\V\nu_{j_1,j_2}) +
    \objphase(\V\nu_{j_2,j_3}) +
    \objphase(\V\nu_{j_3,j_1}) \mod 2\,\pi \, ,
  \label{eq:phase-closure}
\end{align}
where $\arg: \Complexes\mapsto(-\pi,+\pi]$ yields the phase of its argument
and $\objphase(\V\nu) = \arg\Paren{\FT{I}(\V\nu)}$ is the phase of the
Fourier transform of the brightness distribution.

The so-called \vocab{differential visibilities}
\cite{Petrov_et_al-2003-Amber, Duvert_et_al-2015-oifits2} are formed from
the cross-product of the coherent flux and a reference coherence flux:
\begin{equation}
  \diffvis_{j_1,j_2,\ell,m}
  = \Paren[\big]{V_{j_1,j_2,\ell,m}^\Tag{ref}}^{\!\star} \,
  V_{j_1,j_2,\ell,m} \, .
  \label{eq:differential-visibility}
\end{equation}
The purpose of the reference coherence flux $V_{j_1,j_2,\ell,m}^\Tag{ref}$
is to cancel most of the randomness due to the turbulent phase in the
coherent flux $V_{j_1,j_2,\ell,m}$.  To that end, the reference coherence
flux is defined as some spectral average of the coherent flux in the other
spectral channels.  Introducing a matrix $\M W^\Tag{ref}$ of nonnegative
weights, a general expression for the reference flux is:
\begin{equation}
  V_{j_1,j_2,\ell,m}^\Tag{ref} =
  \sum_{\ell' \not= \ell} W_{\ell,\ell'}^\Tag{ref} \, V_{j_1,j_2,\ell',m} \, .
  \label{eq:reference-coherent-flux}
\end{equation}
In the above 2 equations, we used the index notation previously introduced
to indicate clearly that the differential visibilities are measured from
the calibrated coherent fluxes in different spectral channels but for the
same baseline and exposure. The phases of the differential visibilities are
known as the \vocab{differential phases} and are used by some image
reconstruction algorithms to recover $I(\V\theta,\lambda)$ the
multi-spectral brightness distribution of the object.

The power spectra, the bispectra and the differential visibilities are all
insensitive to the random time-varying phase shifts due to the turbulence
and can therefore be temporally integrated to improve their signal to noise
ratio (SNR). As a matter of fact these observables are invariant to a
translation of the object.  Thus its absolute position may not be recovered
from such data. Note that correctly measuring these nonlinear observables
involves removing biases \cite{Dainty_Greenaway-1979-powerspectrum,
Wirnitzer-1985-bispectral_analysis,
Gordon_Buscher-2012-interferometric_noise} whose expressions are not shown
in the above expressions.

\subsection{Complex visibilities and normalization}


In this paper, we will adopt the same convention as in radio-astronomy and
use \vocab{complex visibilities} or just \vocab{visibilities} to mean the
samples of the Fourier transform of the object brightness distribution at
the observed frequencies
\cite{Thompson_et_al-2017-radio_astronomy} whether it is
normalized or not.  In optical interferometry, to cope with the turbulence,
the observables are formed after $V_{j_1,j_2}$ defined in
Eq.~(\ref{eq:measured-visibility}) and thus only depend on the normalized
visibilities which are the Fourier transform of the brightness distribution
given in Eq.~(\ref{eq:normalized-brightness-distribution}) and which are
equal to unity at the zeroth frequency.  In radio-astronomy, it is often the
case that the visibility at the zeroth frequency cannot be measured by the
interferometric array and has to be determined by other means or imposed
during the image reconstruction.  To handle all these cases consistently,
we will hereinafter denote by $I(\V\theta)$ the brightness distribution of
interest and explicitly introduce:
\begin{equation}
  \label{eq:normalization}
  \totalflux
  = \iint I(\V\theta)\,\mathd^2\V\theta
  = \FT{I}(\V0) \, ,
\end{equation}
the total brightness which is also the visibility at the zeroth frequency.
The OI-FITS data format \cite{Pauls_et_al-2005-oifits} used in optical
interferometry assumes normalized visibilities and thus that $\totalflux =
1$.  The proposed new OI-FITS~2 format
\cite{Duvert_et_al-2015-oifits2} adds support for coherent fluxes (among
other things).

%

\section{Imaging from sparse Fourier data}

Image reconstruction from sparse data is a well studied but difficult
problem which is further complicated when dealing with non-linear data such
as the power spectra, the bispectra (or the closure phases) or the
differential visibilities (or the differential phases).  We first consider
reconstruction from the sparse measurements of the Fourier transform of the
brightness distribution.  This is a typical problem for radio-astronomy or
for optical interferometry when a phase reference
\cite{Delplancke_at_al-2003-Prima} is used and thus complex visibilities
are available.  Understanding the principles of image reconstruction
\cite{Titterington-1985-regularization, Demoment-1989-image_reconstruction}
is recommended for the proper use of a given algorithm. Fortunately, most,
if not all, image reconstruction methods follow similar approaches
\cite{Thiebaut-2009-interferometric_imaging,
Thiebaut_Giovannelli-2010-interferometry,
Baron-2016-image_reconstruction_overview} which we will review here.

%

\subsection{Image and complex visibility models}

For practical reasons, the image sought must be a parametric approximation
of $I(\V\theta)$ the (possibly normalized) object brightness distribution.
The image assumed by almost all existing reconstruction algorithms can be
described by a linear expansion like:
\begin{equation}
  \label{eq:linear-image-model}
  I(\V\theta) \approx I_\Tag{lin}(\V\theta)
  = \sum_{n=1}^{N} \Param[n] \, h_n(\V\theta) \, ,
\end{equation}
where $\List{h_n: \Theta \mapsto \Reals}{n=1}{N}$ is a given  basis of
real-valued functions on the field of view $\Theta$ and the coefficients
$\Param \in \Reals^N$ are the image parameters.  The total intensity of this
class of model image is exactly computable:
\begin{equation}
  \label{eq:total-flux}
  \totalflux = \iint I(\V\theta) \, \mathd\V\theta
  \approx \iint I_\Tag{lin}(\V\theta) \, \mathd\V\theta
  = \Inner{\V c}{\Param} \, ,
\end{equation}
where $\V c \in \Reals^N$ has components $c_n = \iint h_n(\V\theta) \,
\mathd\V\theta$.  The Fourier transform of the model image is also exactly
computable:
\begin{equation}
   \FT{I}_\Tag{lin}(\V\nu) = \sum_{n=1}^{N} \Param[n] \, \FT{h}_n(\V\nu) \, ,
\end{equation}
with $\FT{h}_n(\V\nu)$ the Fourier transform of $h_n(\V\theta)$. For the
observed spatial frequencies, the model of the visibility $V_{k}$ is then:
\begin{equation}
  \label{eq:complex-vis-model}
  \FT{I}(\V\nu_k) \approx (\ModelOp\cdot\Param)_k
  = \sum_{n=1}^{N} \ModelOp[k,n] \, \Param[n] \, ,
\end{equation}
where $\ModelOp \in \Complexes^{K \times N}$ is a linear operator whose
coefficients are $\ModelOp[k,n] = \FT{h}_n(\V\nu_k)$ and $K$ is the number
of sampled frequencies.

Although the linear expansion in Eq.~(\ref{eq:linear-image-model}) is
general and can represent specific models such as wavelet-based
decompositions \cite{Starck_et_al-1994-aperture_synthesis,
Baron-2016-image_reconstruction_overview} or a mixture of a pixel-based
image and a parametric model \cite{Kluska_et_al-2014-SPARCO}, a more
typical example is:
\begin{equation}
  \label{eq:grid-image-model}
  I(\V\theta) \approx I_\Tag{grd}(\V\theta)
  = \sum_{n=1}^{N} \Param[n] \, h(\V\theta - \V\theta_n) \, ,
\end{equation}
which is the model in Eq.~(\ref{eq:linear-image-model}) with $h_n(\V\theta)
= h(\V\theta - \V\theta_n)$ where $\V\theta_n$ are angular positions
forming a rectangular equispaced grid. The function $h: \Theta \mapsto
\Reals$ may be thought of as an interpolation function, it is called the
\emph{clean beam} in the \CLEAN method \cite{Hogbom-1974-CLEAN}, the
\emph{neat beam} in the \WIPE method
\cite{Lannes_et_al-1997-Clean_and_Wipe}, a \emph{building block}
\cite{Hofmann_Weigelt-1993-building_blocks} or, more simply, the
\emph{pixel shape} of a conventional image representation.  This family of
models yields $c_n = \iint h(\V\theta) \, \mathd\V\theta$ which is constant
and $\ModelOp[k,n] =
\FT{h}(\V\nu_k)\,\exp(-\mathi\,2\,\pi\,\Inner{\V\theta_n}{\V\nu_k})$.

Finally, we note that most image reconstruction algorithms assume that the
image parameters are just samples of the brightness distribution:
\begin{equation}
  \label{eq:simple-image-model}
  \Param[n] \approx I(\V\theta_n) \quad (\forall n) \, ,
\end{equation}
and compute the visibilities with:
\begin{equation}
  \label{eq:non-uniform-Fourier-transform}
  \ModelOp[k,n] = \exp(-\mathi\,2\,\pi\,\Inner{\V\theta_n}{\V\nu_k})
  \quad(\forall k, \forall n) \,.
\end{equation}
This corresponds to the choice $h_n(\V\theta) = h(\V\theta - \V\theta_n) =
\delta(\V\theta - \V\theta_n)$ in the previous models.  Computing the
visibilities then amounts to performing a \vocab{non-uniform Fourier
transform}. There are fast algorithms, as NUFFT
\cite{Fessler_Sutton-2003-NUFFT} or NFFT \cite{Keiner_et_al-2009-NFFT}, to
compute a good approximation of $\ModelOp\cdot\Param$ by means of the fast
Fourier transform (FFT). Another consequence of this choice, is that
$c_n=1$ ($\forall n$).  Unless otherwise specified, we will assume this
kind of image model in this paper.


The size of the synthesized field of view and of the image pixels have to
be chosen according to the extension of the observed object and to the
resolution of the interferometer.  To avoid biases and inaccurate
approximations caused by the particular image model, the grid spacing
$\Delta\theta$ should be well beyond the limit imposed by the longest
(projected) baseline $b^\Tag{proj}_\Tag{max}$ between any pairs of
interfering telescopes:
\begin{equation}
  \label{eq:diffraction-limit}
  \Delta\theta \ll \frac{\lambda}{2\,b^\Tag{proj}_\Tag{max}} \, .
\end{equation}
Oversampling by a factor of at least 2 is normally used, hence the pixel
size is typically given by: $\Delta\theta \lesssim
\lambda/(4\,b^\Tag{proj}_\Tag{max})$.  To avoid aliasing and image
truncation, the field of view must be chosen large enough and without
forgetting that the reciprocal of the width of the field of view also sets
the sampling step of the spatial frequencies.

\subsection{Obtaining a dirty image}

We now consider a first attempt for image restoration from sparse sampling
of the spatial frequency plane in the most simple case, that is assuming
that complex visibilities have been measured.  In this context, let
$\List{\V\nu_k}{k = 1}{K}$ be the list of sampled frequencies and $\Data
\in \Complexes^K$ be the data ($\Data[k] = V_k$ is the measured complex
visibility at frequency $\V\nu_k$). According to
Eq.~(\ref{eq:complex-vis-model}), $\Data \approx \ModelOp\cdot\Param$ with
$\Param \in \Reals^N$ the image parameters and where $\approx$ accounts for
the model approximations and for the noise. In practice, there are more
pixels in the sought image than measurements ($N \gg 2\,K$) so the image
restoration problem is ill-posed.  However, the relationship between
$\Data$ and $\Param$ is linear and a possible solution that comes to mind
is to estimate $\Param$ by means of the Moore-Penrose pseudo-inverse of
$\ModelOp$ applied to the data $\Data$ (hence neglecting the noise). This
is equivalent to finding the minimal norm image $\BestParam$ such that
$\ModelOp\cdot\BestParam = \Data$ as stated by the constrained optimization
problem:
\begin{equation}
  \label{eq:least-norm-solution}
  \min_{\Param} \Norm{\Param}_2 \quad\text{s.t.}\quad \ModelOp\cdot\Param = \Data
\end{equation}
where $\Norm{\Param}_2 = \sqrt{\!\Inner{\Param}{\Param}}$ is the usual Euclidean
($\ell_2$) norm. The Lagrangian of the above problem is:
\begin{displaymath}
  \mathcal{L}(\Param, \V u)
  = (1/2) \, \Norm{\Param}_2^2 - \Inner{\V u}{\ModelOp\cdot\Param}
  = (1/2) \, \Norm{\Param}_2^2 - \Inner{\ModelOp^\star\cdot\V u}{\Param} \, ,
\end{displaymath}
with $\ModelOp^\star$ the adjoint of $\ModelOp$ and $\V u$ the Lagrange
multipliers associated with the constraints $\ModelOp\cdot\Param = \Data$.
Since $\nabla_{\!\Param}\mathcal{L}(\Param, \V u) = \Param -
\ModelOp^\star\cdot\V u$, the solution takes the form $\BestParam =
\ModelOp^\star\cdot\estim{\V u}$ where $\estim{\V u}$ is such that the
equality constraints hold, that is
$\ModelOp\cdot\ModelOp^\star\cdot\estim{\V u} = \Data$.  Considering the
simple case in Eq.~(\ref{eq:non-uniform-Fourier-transform}) and assuming a
given frequency is only measured once, an approximation of
$\ModelOp\cdot\ModelOp^\star$ is given by:
\begin{displaymath}
  \Paren[\big]{\ModelOp\cdot\ModelOp^\star}_{k,k'}
  = \sum_{n=1}^{N} \exp(-\mathi\,2\,\pi\,\Inner{\V\theta_n}{\V\nu_{k} - \V\nu_{k'}})
  \approx N\,\delta_{k,k'} \, ,
\end{displaymath}
because the sum is equal to $N$ when $\V\nu_{k} = \V\nu_{k'}$ and
approximately equal to zero when $\V\nu_{k} \not= \V\nu_{k'}$ (which is
equivalent to $k \not= k'$ under our assumptions).  Thus $\estim{\V u}
\approx \Data/N$ and:
\begin{equation}
  \label{eq:dirty-map}
  \BestParam \approx (1/N) \,\ModelOp^\star\cdot\Data \, ,
\end{equation}
which is known as the \vocab{dirty image}
\cite{Thompson_et_al-2017-radio_astronomy}.  If the object is a point-like
source (situated at the center of the field of view), all complex
visibilities are equal to $\totalflux$ and the corresponding dirty image is
called the \vocab{dirty beam}.  Examples of a dirty beam and dirty image are
shown in the lower left and right panels of Fig.~\ref{fig:dirty-map};
clearly the general inverse does not provide a good solution to the image
restoration problem.

In radio-astronomy where the complex visibilities can be measured, image
restoration is often treated as the deconvolution of the dirty image by a
point spread function (PSF) which is the dirty beam
\cite{Cornwell_Braun-1989-deconvolution, Giovannelli_Coulais-2005-pos_mix}.
This however assumes a stationary distribution of the noise. The
deconvolution approach is therefore not applicable if one wants to account
for more realistic noise statistics or if complex visibilities are not
measured directly (as is the case in optical interferometry).

\subsection{Improving the reconstruction}
\label{sec:reconstruction-principles}

The dirty image arose from an attempt to solve an ill-posed problem; by
reducing the number of possible solutions one can hope to improve the
situation.  Being a brightness distribution, the image must be nonnegative
everywhere and may be normalized according to Eq.~(\ref{eq:normalization}).
As a consequence, the sought image should be restricted to belong to the
convex set $\Omega \subset \Reals^N$ of nonnegative and normalized images:
\begin{equation}
  \label{eq:feasible-set}
  \Omega = \Brace[\big]{
    \Param \in \Reals^N \Given
    \Param \ge 0,\ \Inner{\V c}{\Param} = \totalflux
  } \, ,
\end{equation}
where $\V c$ is defined in Eq.~(\ref{eq:total-flux}) and the relation
$\Param \ge 0$ taken componentwise expresses the nonnegativity of the pixel
values assuming the simple image model in
Eq.~(\ref{eq:simple-image-model}). Strict constraints such as the image
being nonnegative are effective for image reconstruction from sparse
visibilities but are yet insufficient to select a single image out of all
the ones which fit the data.

Selecting a single image was the purpose of minimizing the Euclidean norm
in Problem~(\ref{eq:least-norm-solution}). By Parseval's theorem, the
Euclidean norm of the image is also that of its Fourier transform, thus the
Fourier transform of the dirty image is equal to the measured visibilities at
the sampled frequencies and equal to zero elsewhere. This explains the
ripples and the poor quality of the dirty image. Intuitively, a more
appealing solution would be an image whose Fourier spectrum interpolates
the sparse data in a smoother way than that of the dirty image. In other
words, a simpler or more regular image than the dirty image is to be
preferred. Accounting for the strict constraints implemented by $\Omega$,
we are then inclined to reformulate Problem~(\ref{eq:least-norm-solution})
as:
\begin{equation}
  \label{eq:least-prior-solution}
  \min_{\Param \in \Omega} \Fprior(\Param)
  \quad\text{s.t.}\quad
  \ModelOp\cdot\Param = \Data \, ,
\end{equation}
where $\Data$ denote measured complex visibilities as before and minimizing
$\Fprior:\Reals^N\mapsto\Reals$ is intended to favor a regular image in
agreement with our prior beliefs. Clearly the solution will depend on the
type of priors implemented by $\Fprior$ and known as the
\vocab{regularization}.  For instance, taking $\Fprior(\Param) =
\Norm{\Param}_2$ and $\Omega = \Reals^N$ yields the dirty image.


Before discussing suitable choices for the regularization, there are
however other issues to consider related to the equality constraint $\Data
= \ModelOp\cdot\Param$ in (\ref{eq:least-prior-solution}).  First,
visibilities may not be directly measured and the actual data $\Data$ may
comprise nonlinear quantities like the power spectrum and the bispectrum
(or the closure phase).  A more general form, say
$\ModelFn(\Param,\Hyper)$, must be introduced to model such nonlinear data
(here $\Hyper$ denotes any other parameters besides $\Param$, like the
pixel size $\Delta\theta$, which impact the model).  Second, real data are
corrupted by noise and the equality constraint should be replaced by
something like $\Data \approx \ModelFn(\Param,\Hyper)$.  To be more
specific, this can be expressed by the following \vocab{direct model} (also
called \emph{forward model}
\cite{Gerwe_et_al-2013-intensity_interferometry_imaging}) of the data:
\begin{equation}
\label{eq:direct-model}
  \Data = \ModelFn(\Param,\Hyper) + \Noise \, ,
\end{equation}
where $\Noise$ is a random perturbation term due to the noise.  Because of
this random perturbation, an exact fit of the measurements is not only
unexpected but undesirable (we do not want to fit the noise).  In fact, any
image should be considered as acceptable provided that the corresponding
model data differ from the actual data by amounts consistent with the noise
level.  In order to judge quantitatively whether an image $\Param$ is
statistically consistent with the measurements, some numerical criterion,
say $\Fdata: \Reals^N \mapsto \Reals$, is needed.  By convention, the
smaller $\Fdata(\Param)$ the closer the model data are to the measurements,
so $\Fdata(\Param)$ can be thought of as a \emph{distance} between the
model and the data. In other words and using the metric $\Fdata(\Param)$,
an image should be assumed to be compatible with the data whenever
$\Fdata(\Param)$ is below some threshold, say:
\begin{equation}
  \label{eq:discrepancy}
  \Fdata(\Param) \le \eta  \, .
\end{equation}
According to this discrepancy criterion, the image reconstruction problem
becomes:
\begin{equation}
  \label{eq:constrained-problem}
  \min_{\Param \in \Omega} \Fprior(\Param)
  \quad \text{s.t.}\quad
  \Fdata(\Param) \le \eta \,,
\end{equation}
which formally expresses that we search for ``\emph{the most regular image
which is compatible with the observations and for which the strict
constraints hold}.'' The usual way to solve the constrained
problem~(\ref{eq:constrained-problem}) is to use the associated Lagrangian:
\begin{equation}
  \label{eq:Lagrangian}
  \mathcal{L}(\Param, \alpha) = \Fprior(\Param) + \alpha\, \Fdata(\Param) \, ,
\end{equation}
with $\alpha \ge 0$ the Lagrange multiplier related to the constraint
$\Fdata(\Param) \le \eta$.  Technically, the multiplier $\alpha$ must be
nonnegative because the constraint is an inequality
\cite{Nocedal_Wright-2006-numerical_optimization}.  If $\alpha = 0$, the
constraint has no effects (it is said to be \emph{inactive}) which, in our
case, means that the data are completely ignored in determining the
resulting image.  This is only possible if the threshold $\eta$ is
sufficiently high that the inequality $\Fdata(\Param) \le \eta$ holds at
the minimum of $\Fprior(\Param)$.  In other words, our priors are
sufficient to determine an acceptable image regardless of the data. This is
not generally the case and we want to take the data into account (the
constraint must be \emph{active}) which implies that $\alpha > 0$.  It
turns out that this also amounts to assuming that the constraint be
$\Fdata(\Param) = \eta$ at the solution
\cite{Nocedal_Wright-2006-numerical_optimization}.

To summarize, image reconstruction is recast as a constrained optimization
problem where the solution $\BestParam$ is given by:
\begin{equation}
  \label{eq:constrained-solution}
  \BestParam = \argmin_{\Param \in \Omega} \bigl\{
     \mathcal{L}(\Param, \alpha) =
     \Fprior(\Param) + \alpha \, \Fdata(\Param)
  \bigr\} \, ,
\end{equation}
with $\alpha > 0$ chosen so that $\Fdata(\BestParam) = \eta$ holds.
Since $\alpha > 0$, taking $\mu = 1/\alpha$ yields the following equivalent
formulation:
\begin{equation}
  \label{eq:alternative-solution}
  \BestParam = \argmin_{\Param \in \Omega} \bigl\{
     f(\Param, \mu) =
     \Fdata(\Param) + \mu\,\Fprior(\Param)
  \bigr\} \, ,
\end{equation}
where $\mu > 0$ is tuned so that the constraint holds.  The solution
$\BestParam$ depends on the feasible set $\Omega$, on the penalties
$\Fdata$ and $\Fprior$ and on the value of a so-called
\vocab{hyper-parameter} (here $\eta$, $\alpha$ or $\mu$).  The
hyper-parameter $\mu > 0$ (resp.\ $\alpha > 0$) can be seen as a relative
weight which sets the compromise between fitting the data (and the noise)
and obeying the priors.

Depending on the image reconstruction algorithm being considered, one of
the equivalent problems (\ref{eq:constrained-problem}),
(\ref{eq:constrained-solution}) or (\ref{eq:alternative-solution}) is
solved and the hyper-parameter (equivalently $\eta$, $\alpha$ or $\mu$) is
explicitly required or automatically tuned by the method.  Algorithms
mostly depend on the choice for $\Fdata$ and $\Fprior$ --- the following
section provides some hints to define these penalties.

\subsection{Bayesian inference}

Another approach to derive a general expression for the solution of the
image reconstruction problem is to consider that all information is
expressed in terms of probabilities.  In this Bayesian framework, the best
image given the data is the \emph{maximum a posteriori} (MAP) solution
which has the maximum posterior probability given the data and, possibly,
some supplementary parameters $\Hyper$:
\begin{equation}
    \BestParam_\Tag{MAP} = \argmax_{\Param} \PDF(\Param \given \Data, \Hyper)
\end{equation}
where $\Data$ represents all the available data.  The supplementary
parameters $\Hyper$ represent anything other than the data that influences
the expression of the probabilities (the covariance of the noise,
\emph{etc.}).  Bayes' rule gives two equivalent expressions for the joint
probability of $\Param$ and $\Data$ (knowing $\Hyper$):
\begin{align*}
  \PDF(\Param, \Data \given \Hyper)
  &= \PDF(\Param \given \Hyper) \, \PDF(\Data \given \Param, \Hyper) \notag \\
  &= \PDF(\Data \given \Hyper) \, \PDF(\Param \given \Data, \Hyper) \, ,
\end{align*}
from which can be deduced that:
\begin{equation}
  \label{eq:map-def}
  \BestParam_\Tag{MAP} = \argmax_{\Param} \frac{
    \PDF(\Data \given \Param, \Hyper) \, \PDF(\Param \given \Hyper)
  }{
    \PDF(\Data \given \Hyper)
  } \, ,
\end{equation}
where $\PDF(\Data \given \Param, \Hyper)$ is the \vocab{likelihood} of the
data, $\PDF(\Param \given \Hyper)$ is the \emph{a priori} distribution of
$\Param$ and the denominator $\PDF(\Data \given \Hyper)$ is called the
\vocab{evidence}. Discarding the denominator (which does not depend on the
unknowns $\Param$) and taking the co-logarithm of the probabilities (which
converts the product into a sum and the maximum into a minimum) yields:
\begin{equation}
  \BestParam_\Tag{MAP} = \argmin_{\Param \in \Omega} \Brace[\big]{
    -\log\PDF(\Data \given \Param, \Hyper) - \log\PDF(\Param \given \Hyper)
  } \, ,
\end{equation}
where the feasible set $\Omega = \Brace{\Param \in \Reals^N \Given
\PDF(\Param \given \Hyper) > 0}$ is introduced so that taking the logarithm
causes no problems.  Minimizing the first term, $-\log\PDF(\Data \given
\Param, \Hyper)$, amounts to maximizing the likelihood of the data and thus
the agreement of the model with the data.  This is the same objective as
minimizing $\Fdata(\Param)$. Similarly, minimizing the second term,
$-\log\PDF(\Param \given \Hyper)$, enforces the agreement with the priors;
exactly the purpose of $\Fprior(\Param)$. The analogy with the previous
section is evident and may be formalized by taking in
Eq.~(\ref{eq:alternative-solution}):
\begin{align}
  \label{eq:bayesian-likelihood}
  \Fdata(\Param) &= -c_0\,\log\PDF(\Data \given \Param, \Hyper) + c_1\,, \\
  \label{eq:bayesian-regularization}
  \mu\,\Fprior(\Param) &= -c_0\,\log\PDF(\Param \given \Hyper) + c_2\,,
\end{align}
where the factor $c_0 > 0$ and the additive constants $c_1$ and $c_2$ are
irrelevant and can be chosen so as to simplify the resulting expressions.
Clearly the hyper-parameter $\mu$ is part of $\Hyper$. Bayesian inference
can be invoked to justify the heuristic approach of the previous section
but also to derive expressions for the terms of the criterion to optimize.

Without any priors, the solution would be:
\begin{equation}
  \BestParam_\Tag{ML}
  = \argmin_{\Param \in \Omega} \Fdata(\Param)
  = \argmax_{\Param} \PDF(\Data \given \Param, \Hyper) \, ,
\end{equation}
which is nothing else but the maximum likelihood (ML) solution.  In many
estimation problems the maximum likelihood may provide an estimator with
excellent properties, but for solving an ill-posed or ill-conditioned
problem as our image restoration problem, taking into account the priors is
crucial and $\BestParam_\Tag{MAP}$ will be superior to
$\BestParam_\Tag{ML}$ \cite{Titterington-1985-regularization,
Tarantola-1987-inverse_problem_theory, Demoment-1989-image_reconstruction,
Thiebaut-2013-Frejus}.

\section{Likelihood of the data}

Following the Bayesian prescription, Eq.~(\ref{eq:bayesian-likelihood})
states that the correct way to express $\Fdata(\Param)$ is to take the
co-logarithm of the likelihood of the data $\Data$ knowing the image
parameters $\Param$ and perhaps some other parameters $\Hyper$.  We recall
here the direct model given in  Eq.~(\ref{eq:direct-model}):
\begin{displaymath}
  \Data = \ModelFn(\Param,\Hyper) + \Noise \, ,
\end{displaymath}
with $\ModelFn(\Param,\Hyper)$ a parametric model and $\Noise$ a stochastic
term.  If we were able to repeat the same observations (under the same
conditions, with the same instrument and for the same object) many times,
we could average these data and so that the nuisance term $\Noise$ vanishes
in the averaged data.  In other words, the deterministic part of the direct
model is the expectation of the data given the parameters $\Param$ and
$\Hyper$:
\begin{equation}
  \label{eq:data-model-definition}
  \Expect\Brace{\Data\Given\Param,\Hyper} = \ModelFn(\Param,\Hyper) \, ,
\end{equation}
where $\Expect$ denotes expectation.  From this it immediately follows that
the noise is \emph{conditionally centered} in the sense that:
\begin{equation}
  \label{eq:centered-noise}
  \Expect\Brace{\Noise\Given\Param,\Hyper} = \V0 \, .
\end{equation}
From these equations, it is clear that the distribution of the data $\Data$
knowing $\Param$ and $\Hyper$ is simply the distribution of the noise
$\Noise$ (also knowing $\Param$ and $\Hyper$) to which a given bias
$\ModelFn(\Param,\Hyper)$ has been added. Note that these definitions and
their consequences are very general: the model $\ModelFn(\Param,\Hyper)$
may account for any kind of data and the noise $\Noise$ may have any
distribution provided it is conditionally centered as stated by
Eq.(\ref{eq:centered-noise}).


\subsection{Gaussian noise and linear model}

Because the noise term $\Noise$ results from different contributions
(detector noise, photon noise, \etc.), its actual distribution may be quite
complex.  A flexible approximation which works well in practice is to
assume that the noise, knowing $\Param$ and $\Hyper$, follows a centered
Gaussian distribution. Then applying Eq.~(\ref{eq:bayesian-likelihood})
with $c_0 = 2$ (a usual choice) and $c_1$ chosen to discard all additive
constants yields:
\begin{equation}
  \label{eq:fdata-Gaussian}
  \Fdata(\Param) =
  \QuadTerm{\M W}{(\Data - \ModelFn(\Param,\Hyper))} \,,
\end{equation}
where the weighting matrix $\M W = \Cov(\Noise\Given\Param,\Hyper)^{-1}$ is
the inverse of the covariance of the noise.  The above
expression is the usual $\chi^2$ (chi-squared) of the data. There is a
slight issue because, in interferometry, we may be dealing with complex
valued data. Without loss of generality and since complex numbers are just
pairs of reals, complex valued vectors (such as $\Data$, $\Noise$ and
$\ModelFn(\Param,\Hyper)$) can be \emph{flattened} into ordinary real
vectors (with doubled size) to use standard linear algebra notation and
define the covariance matrix as $\Cov(\Noise\Given\Param,\Hyper) =
\Expect\Brace{\Noise\cdot\Noise\T\Given\Param,\Hyper}$.  This is what is
assumed in Eq.~(\ref{eq:fdata-Gaussian}).

When the complex visibilities are directly measured, the model given by
Eq.~(\ref{eq:complex-vis-model}) is linear and the likelihood
$\Fdata(\Param)$ in Eq.~(\ref{eq:fdata-Gaussian}) is quadratic (by
construction) and \emph{convex} because its Hessian,
$\nabla^2\Fdata(\Param) = \ModelOp\T\cdot\M W\cdot\ModelOp$, is positive
semi-definite.  Having $\Fdata(\Param)$ convex in $\Param$ is particularly
suitable for optimization because it generally guarantees that there exists
a unique minimum which can be easily found by means of the most simple
methods such as gradient-based optimization algorithms described in Section
\emph{Optimization Strategies}
\cite{Nocedal_Wright-2006-numerical_optimization,
Boyd_Vandenberghe-2009-convex_optimization}.  Conversely, non-convex
functions are more difficult to minimize especially when they are
\emph{multi-modal}, that is to say when they have multiple minima. The most
simple example of a quadratic convex likelihood is given by:
\begin{equation}
  \label{eq:simple-data-criterion}
  \Fdata(\Param) = \sum\nolimits_k w_k \,
  \bigl\vert (\ModelOp\cdot\Param)_k - V_k\bigr\vert^2 \, ,
\end{equation}
where $w_k \ge 0$ are statistical weights.  The above expression which has
been largely used in radio-astronomy \cite{Ables-1974-MEM,
Skilling_Bryan-1984-maximum_entropy,
Cornwell_Evans-1985-simple_maximum_entropy,
Narayan_Nityananda-1986-maximum_entropy_review} assumes that the data are
independent and that the real and imaginary parts of a measured visibility,
say $V_k$, are independent and have the same variance equal to the
reciprocal of $w_k$.  The properties of such a distribution are reviewed in
Goodman's book \cite{Goodman-2015-statistical_optics}.

Real data may however have different statistics.  For instance, the OI-FITS
 \cite{Pauls_et_al-2005-oifits} exchange file format for optical
interferometric data assumes that the errors of complex data (complex
visibilities or bispectra) along the complex vector and perpendicular to
the complex vector are independent but not necessarily of the same variance
(the standard deviations of the amplitude and the phase are provided by the
OI-FITS format). It has however been empirically observed that the errors
for bispectrum data have a \emph{banana-shaped} distribution
\cite{Thiebaut-2013-Frejus} which is compatible with assuming that the
amplitude and phase of the bispectrum are independent. Meimon et al.\
\cite{Meimon_et_al-2005-convex_approximation} have proposed quadratic
convex approximations of the co-log-likelihood of complex data independent
in phase and amplitude and have shown that their so-called \emph{local
approximation} yields the best results, notably when dealing with low
signal to noise data.  This approximation amounts to the model assumed in
OI-FITS \cite{Pauls_et_al-2005-oifits}:
\begin{equation}
  \label{eq:fdata-local-approx}
  \Fdata(\Param) = \sum_k \Brace*{
    \frac{\Re\Paren{\Noise[k]\,\mathe^{-\mathi\,\varphi_k}}^2}
         {\Var(\rho_k)}
    + \frac{\Im\Paren{\Noise[k]\,\mathe^{-\mathi\,\varphi_k}}^2}
           {\rho_k^2\,\Var(\varphi_k)}
  } \, ,
\end{equation}
where $\rho_k$ and $\varphi_k$ are the amplitude and the phase of the
$k$-th complex datum while $\Noise = \Data - \ModelFn(\Param,\Hyper)$
denotes the complex residuals introduced in Eq.~(\ref{eq:direct-model}).
Equation~(\ref{eq:simple-data-criterion}) with $w_k = 1/\Var(\rho_k)$ is
retrieved when $\Var(\varphi_k) = \Var(\rho_k)/\rho_k^2$ which is likely to
result owing to the measurement process
\cite{Soulez_et_al-2014-optimal_fringe_tracking_Montreal}.

\subsection{Nonlinear data}

Dealing with measured complex visibilities is only the simplest case.
In order to account for heterogeneous data consisting of the various
interferometric observables previously described, the likelihood term
may be written:
\begin{equation}
  \label{eq:fdata-heterogeneous}
  \Fdata(\Param) = f_\VisibilityTag(\Param) + f_\PowerspectrumTag(\Param) +
    f_\BispectrumTag(\Param) + f_\ClosurePhaseTag(\Param) \, ,
\end{equation}
where the different terms are for measured complex visibilities, power
spectra, bispectra and closure phases respectively.  Such an expression
assumes that the measurements taken into account by different terms are
independent. The bispectra and closure phases which are obviously not
independent should not be considered at the same time.  A revised OI-FITS
format has been proposed to provide, among other things, the correlations
between the measurements \cite{Duvert_et_al-2015-oifits2} but, to our
knowledge, this information is not yet used by existing image
reconstruction algorithms.

For power spectrum data, most algorithms assume independent Gaussian
measurements and take:
\begin{equation}
  \label{eq:powerspectrum-penalty}
  f_\PowerspectrumTag(\Param) = \sum_{k} w^\PowerspectrumTag_{k}
  \,\Paren*{
    S_{k} - \Abs{\FT{I}(\V\nu_{k})}^2
  }^2 \, ,
\end{equation}
where $S_k$ is the power spectrum measured at frequency $\V\nu_k$,
$w^\PowerspectrumTag_{k} = 1/\Var(S_k)$ and $\FT{I}(\V\nu_{k}) =
(\ModelOp\cdot\Param)_{k}$ is the model of the visibility.  Note that there
are certainly other distributions more appropriate than this one
\cite{Goodman-2015-statistical_optics}.

For bispectrum data, the OI-FITS data model would suggest approximating
$f_\BispectrumTag(\Param)$ following Eq.~(\ref{eq:fdata-local-approx}), but
a common expression is:
\begin{equation}
  \label{eq:bispectrum-penalty}
  f_\BispectrumTag(\Param) = \sum_{k_1,k_2} w^\BispectrumTag_{k_1,k_2}
  \,\Abs*{
    B_{k_1,k_2} -
    \FT{I}(\V\nu_{k_1})\,
    \FT{I}(\V\nu_{k_2})\,
    \FT{I}^\star(\V\nu_{k_1} + \V\nu_{k_2})
  }^2 \, ,
\end{equation}
where $B_{k_1,k_2}$ is the measured bispectrum for frequencies
$\V\nu_{k_1}$ and $\V\nu_{k_2}$.  This
expression is consistent if, as for Eq.~(\ref{eq:simple-data-criterion}),
the real and imaginary parts of $B_{k_1,k_2}$ are independent Gaussian
variables with the same variance equal to the reciprocal of the weights
$w^\BispectrumTag_{k_1,k_2}$. However, the building-block
\cite{Hofmann_Weigelt-1993-building_blocks} and \IRBIS
\cite{Hofmann_et_al-2014-IRBIS} algorithms assume a different form
of the weights which attempts to compensate for the uneven sampling of the
frequency plane.

For phase only data, it is appropriate to assume a von Mises distribution
\cite{Mardia-1975-directional_data_statistics} for the wrapped phase (see
Appendix \emph{Penalty for Angular Data}). For closure phase data, defined
in Eq.~(\ref{eq:phase-closure}), this leads to:
\begin{align}
  f_\ClosurePhaseTag(\Param)
  = 2\,\sum_{k_1,k_2} \kappa^\ClosurePhaseTag_{k_1,k_2}\,
  \bigl[
    1 - \cos\bigl(&
      \closurephase_{k_1,k2}\textit{}
      - \objphase(\V\nu_{k_1})
      - \objphase(\V\nu_{k2})
     \notag \\[-2ex]
   &+ \objphase(\V\nu_{k_1}+\V\nu_{k_2})
     \bigr)
   \bigr]
   \, ,
  \label{eq:closure-penalty-von-Mises}
\end{align}
where $\kappa^\ClosurePhaseTag_{k_1,k_2} > 0$ can be computed from the
angular variance $\Var(\closurephase_{k_1,k_2})$ of the closure phase and
using Eq.~(\ref{eq:angular-variance}) derived in the Appendix.  An
expression similar to Eq.~(\ref{eq:closure-penalty-von-Mises}) can be used
to fit the differential phases \cite{Schutz_et_al-2014-Painter}.


\section{Regularizations}
\label{sec:regularization}


While the Bayesian formalism rather straightforwardly provides a suitable
definition of the likelihood term, the situation is not as clear for the
regularization term which enforces the priors.  Indeed, in most realistic
cases, the prior probability cannot be objectively inferred and one has to
take a more practical point of view (for instance the one introduced in
Section \emph{Improving the Reconstruction}). When selecting a specific
regularization, one has to pay great attention to: (i) the ability of the
priors to help interpolate the missing Fourier samples; and (ii) the type
of bias that is imposed in the restored image.  The \vocab{default
solution} $\Param^\Tag{def}$ which is retrieved when there are no available
data (or when $\mu \rightarrow \infty$ in
Problem~\ref{eq:alternative-solution}) may be used to determine the type of
bias imposed by the regularization.  The default image is given by:
\begin{equation}
   \label{eq:default-solution}
   \Param^\Tag{def} \in \argmin_{\Param \in \Omega} \Fprior(\Param) \, ,
\end{equation}
being aware that the solution of this problem may not be unique.

Existing software exploit various regularizations and a review of their
relative merits is available \cite{Renard_et_al-2011-regularization}.  To
avoid confusion when discussing the many possible regularizations, a simple
classification is needed. 
First, a regularization function can be \vocab{separable} (\ie, the sum of
univariate functions) or not.  Non-separable regularizations usually impose
some sort of smoothness for the solution while separable regularizations
implement a \emph{white prior} and do not impose smoothness (in the image
domain).  A second property to consider is whether the regularization is
quadratic or not.  Quadratic functions are easier to optimize but
non-quadratic ones may have better properties such as suppression of
ripples or a preference for sparse solutions.

\subsection{Quadratic regularizations}
\label{sec:quadratic-regularization}

\begin{figure*}[t]
  \centering
  \includegraphics[height=110mm]{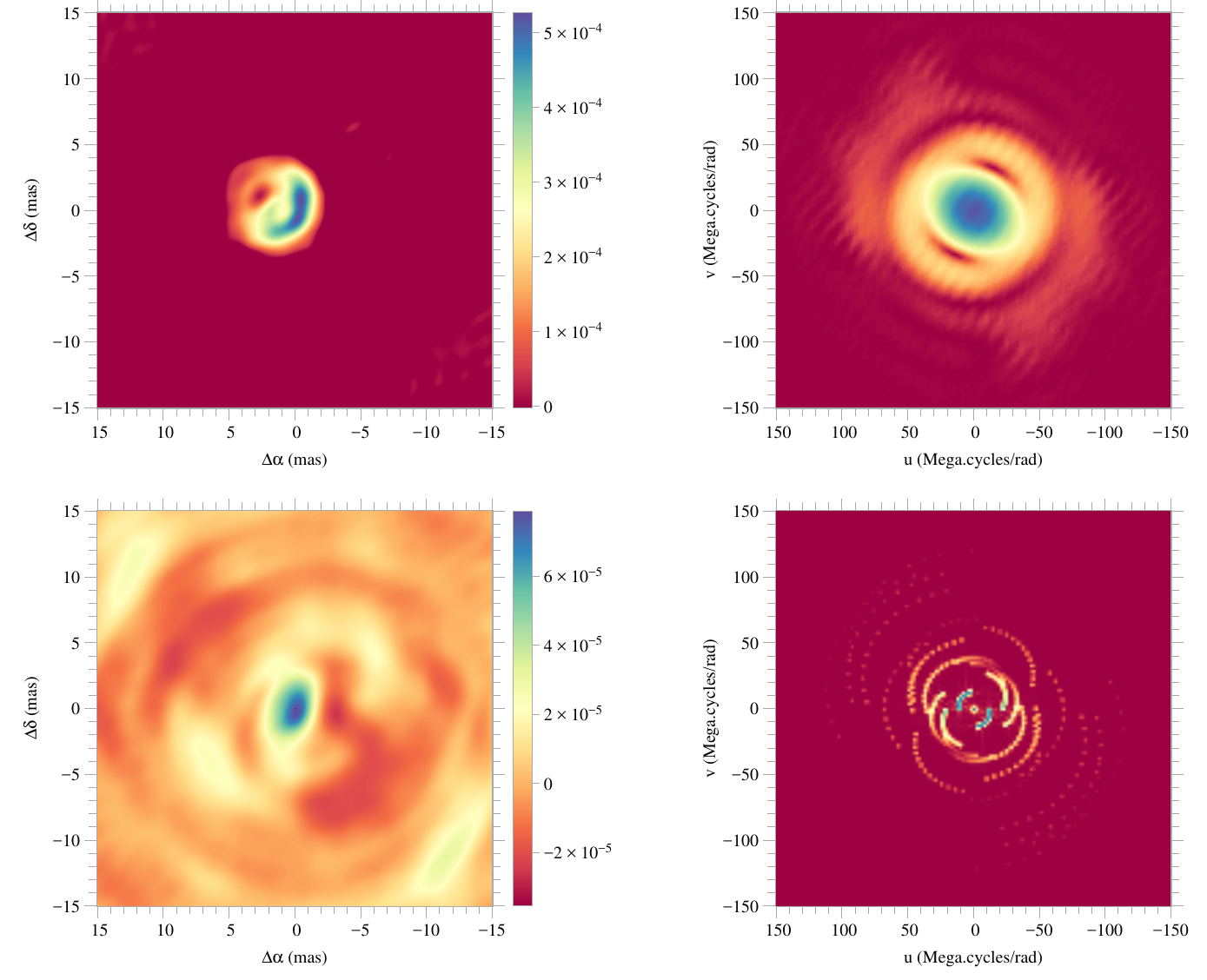}

  \caption{Importance of imposing the positivity.  Top: reconstruction with
  the positivity constraint. Bottom: unconstrained reconstruction. Left:
  restored images.  Right: power spectrum.  All reconstructions were
  performed with MiRA on the simulated data of the \emph{2004
  Interferometric Beauty Contest} and with a \emph{compactness constraint}
  such that the default solution has a fairly large FWHM of 50\,mas.}
  \label{fig:positivity?}

\end{figure*}

Although it does not impose positivity, the well known \emph{Wiener filter}
yields the MAP solution given by Eq.~(\ref{eq:map-def}) when assuming
Gaussian distributions (for the prior and the likelihood) with a linear
model as in Eq.~(\ref{eq:complex-vis-model}).  Assuming a Gaussian prior
leads to a quadratic prior whose general expression is:
\begin{equation}
  \label{eq:quadratic-prior}
  \Fprior(\Param) = \Norm{\M D\cdot\Param - \V a}_2^2 \, ,
\end{equation}
with $\M D$ a given linear operator transforming the space of image
parameters to some other space and $\V a$ a given element of this latter
space.  Quadratic regularizations are very popular because they are easy to
optimize and are quite versatile for imposing various types of prior
knowledge.  For instance, in many image restoration problems, the
smoothness of the solution is favored by taking $\V a = \V0$ and $\M D$ a
finite difference operator whose output collects the differences between
neighboring pixels.  Taking $\M D$ equal to the identity and $\V a$ equal
to zero yields classical Tikhonov regularization
\cite{Tikhonov_Arsenin-1977-ill_posed_problems}.

Without the positivity constraint, Tikhonov regularization yields the
dirty image of Eq.~(\ref{eq:dirty-map}).  For this reason, quadratic
regularizations have long been considered as being badly adapted to
interferometric image reconstruction
\cite{Narayan_Nityananda-1986-maximum_entropy_review}.  The following
simple quadratic regularization has however proven
\cite{leBesnerais_et_al-2008-interferometry} to be very effective in
this context:
\begin{equation}
  \label{eq:quadratic-compactness}
  \Fprior(\Param) = \sum_n q_n \, \Param[n]^2 = \Norm{\Param}_{\M Q}^2 \, ,
\end{equation}
where the components of $\V q \in \Reals^N_+$ are nonnegative weights and
$\M Q = \Diag(\V q)$ is the diagonal matrix whose diagonal elements are
given by $\V q$.  This regularization amounts to taking $\V a =0$ and $\M
D\T\cdot\M D = \M Q = \Diag(\V q)$ in Eq.~(\ref{eq:quadratic-prior}).  As
this regularization is separable, the default solution can be computed
following Appendix \emph{Separable Regularizations} (and provided $\V q >
0$):
\begin{equation}
  \label{eq:default-quadratic-compactness}
  \Param^\Tag{def}
  = \argmin_{\Param \in \Omega} \Norm{\Param}_{\M Q}^2
  = \frac{
    \totalflux\,\V c/\V q
  }{
    \Inner{\V c}{\V c/\V q}
  } \, ,
\end{equation}
where the division in $\V c/\V q$ is performed componentwise. Hence the
shape of the default solution is simply given by $\V c/\V q$.  This
behavior may be exploited to impose a \vocab{compactness prior} by having
$q_n$ be an increasing function of $\Norm{\V\theta_n}$ the angular distance
of the $n$-th pixel to the center of the field of view and thus favor the
concentration of the flux in the central part of the image
\cite{leBesnerais_et_al-2008-interferometry,
Renard_et_al-2011-regularization}.  Figure~\ref{fig:positivity?} shows that
such a regularization is suitable for image reconstruction from optical
interferometric data provided strict positivity is imposed.  Comparing the
power spectra displayed by this figure with the frequency coverage in
Fig.~\ref{fig:dirty-map}, it is clear that the unmeasured frequencies have
been set to zero in the unconstrained solution while they are smoothly
interpolated when the constraint is imposed.  The reason for this ability
is that imposing the positivity plays the role of a floating support
constraint which results in a smoothing in the frequency domain.  This
example demonstrates how critical the positivity constraint is for
interferometric image reconstruction.  Note that the unconstrained solution
in Fig.~\ref{fig:positivity?} is similar but not exactly like the dirty
image in Fig.~\ref{fig:dirty-map} because it was obtained from simulated
power spectra and closure phases (not from complex visibilities).

The Wiener filter is usually written as a separable filter in the frequency
space, then it amounts to implementing a regularization on the power
spectrum of the image of the form:
\begin{equation}
  \label{eq:spectral-density-regularization}
  \Fprior(\Param) = \sum_k \FT{q}_k \, \Abs{\FT{\Param[]}_k}^2 \, ,
\end{equation}
where $\FT{\Param} = \M F \cdot \Param$ is the Discrete Fourier Transform
(DFT) of $\Param$, $\M F$ is the DFT operator and $\FT{q}_k \ge 0$ are
nonnegative spectral weights.  Taking $\V D = \Diag(\FT{\V q}^{1/2})\cdot\M
F$ and $\V a = \V0$, yields the general expression in
Eq.~(\ref{eq:quadratic-prior}). In the case of the Wiener filter,
$\FT{q}_k$ is the reciprocal of the expected spectral density
$\Expect\Brace{\Abs{\FT{\Param[]}_k}^2}$ of the image and such a
regularization has been successfully used used for blind or myopic
deconvolution \cite{Conan_Mugnier_et_al-1998-myopic-deconvolution}.  In the
context of interferometry, taking $\FT{q}_k = 0$ below a chosen frequency
and $\FT{q}_k = 1$ above that frequency has been used to loosely impose a
cutoff frequency in the restored image
\cite{Lannes_et_al-1997-Clean_and_Wipe}.  Quadratic regularization of the
power spectrum $\Abs{\FT{\Param[]}_k}^2$ has also been proposed
\cite{Gerwe_et_al-2013-intensity_interferometry_imaging} which results in a
quartic penalty in $\Param$.

Without the strict constraints implemented by $\Omega$, the default
solutions favored by a quadratic regularization form a subset:
\begin{equation}
   \Param^\Tag{def} \in \M D^\dagger\cdot\V a + \ker(\M D) \, ,
\end{equation}
where $\ker(\M D)$ and $\M D^\dagger$ are the null-space and the
pseudo-inverse of $\M D$.  Owing to the crucial role of the constraints
implemented by  $\Omega$, notably the positivity, this default solution is
merely of academic interest.

\subsection{Improved smoothness}
\label{sec:non-quadratic-smoothness}

Even though the assumption of smoothness for an extended object is
generally justified, it seems a bit counterproductive to impose this as a
strong prior when we worked so hard to measure the high frequencies of the
object. Moreover imposing smoothness via a quadratic regularization yields
artifacts in the form of ripples around sharp structures such as point-like
sources or straight edges.  To avoid these artifacts, the penalty
implemented by the regularization must be less demanding than a quadratic
cost for large differences between nearby pixels.  This leads to the
introduction of a non-quadratic smoothness regularization given by:
\begin{equation}
  \label{eq:smoothness}
  \Fprior(\Param)
  = \sum\nolimits_{n} \zeta(\M D_n \cdot \Param)
\end{equation}
with $\M D_n$ a linear operator such that $\M D_n \cdot \Param$
approximates the spatial gradient $\nabla\!I(\V\theta_n)$ of the image
$\Param$ around the positions $\V\theta_n$ of the grid of pixels and
$\zeta: \Reals^2 \mapsto \Reals$ is a non-quadratic measure of the length
of its argument.  Usually $\M D_n$ is implemented by means of finite
differences. The actual effects of this family of regularizations is
determined by the function $\zeta$.  Note that quadratic smoothness is
imposed if $\zeta(\V v) = \Norm{\V v}_2^2$ where $\Norm{\V v}_2$ denotes
the usual Euclidean ($\ell_2$) norm.

A very popular example of this family is \vocab{total
variation} (TV) \cite{Rudin_et_al-1992-total_variation} which amounts to
taking:
\begin{equation}
  \label{eq:tv-loss}
  \zeta_\Tag{TV}(\V v) = \Norm{\V v}_2 \, ,
\end{equation}
that is simply the Euclidean norm of $\V v = \M D_n \cdot \Param$ (though
not squared).  Such a regularization promotes the sparsity of the spatial
gradients of the reconstructed image (see Section \emph{Sparsity Promoting
Priors}), that is to say an image where most gradients are zero.  Using
total variation therefore favors piecewise flat images which produces
undesirable \emph{cartoon-like artifacts} (see top-left panel of
Fig.~\ref{fig:regularization-levels}).

\vocab{Edge preserving smoothness}
\cite{Charbonnier_et_al-1997-edge_preserving} is able to preserve sharp
structures while avoiding the cartoon effect of TV.  It involves
designing $\zeta(\V v)$ to have the following asymptotic behavior:
\begin{align}
  \zeta(\V v) \approx \begin{cases}
    \displaystyle O\Paren[\big]{\Norm{\V v}_2^2}
    & \text{for $\Norm{\V v}_2 \ll \tau$,} \\
    \displaystyle O\Paren[\big]{\Norm{\V v}_2}
    & \text{for $\Norm{\V v}_2 \gg \tau$,} \\
  \end{cases}
\end{align}
where $\tau > 0$ is some chosen threshold which sets the transition between
the quadratic ($\ell_2$) and the linear ($\ell_1$) behavior of $\zeta(\V
v)$ which is called an $\ell_1$--$\ell_2$ norm of $\V v$.  Note that using
an $\ell_1$--$\ell_2$ norm requires tuning of two hyper-parameters: $\tau$
and $\mu$.  There are many possible ways to implement an $\ell_1$--$\ell_2$
norm.  For instance the hyperbolic function:
\begin{equation}
  \label{eq:hyperbolic-loss}
  \zeta_\Tag{hyperbolic}(\V v)  = \tau\,\sqrt{\Norm{\V v}_2^2 + \tau^2} - \tau^2 \, ,
\end{equation}
the Huber semi-norm:
\begin{equation}
  \label{eq:Huber-loss}
  \zeta_\Tag{Huber}(\V v) = \begin{cases}
    (1/2)\,\Norm{\V v}_2^2 & \text{if $\Norm{\V v}_2 \le \tau$,} \\
    \Norm{\V v}_2\,\tau - \tau^2/2 & \text{otherwise,} \\
  \end{cases}
\end{equation}
or the \emph{fair} loss function \cite{Rey-1983-robust_estimation} used in
\WISARD \cite{Meimon_et_al-2005-weak_phase_imaging}:
\begin{equation}
  \label{eq:fair-loss}
  \zeta_\Tag{fair}(\V v)
  = \Norm{\V v}_2\,\tau - \log\Paren[\big]{1 + \Norm{\V v}_2/\tau}\,\tau^2 \, .
\end{equation}

%
%

\subsection{Weighted $\ell_p$ norm}
\label{sec:Lp-norm}

Generalizing the weighted quadratic norm of
Eq.~(\ref{eq:quadratic-compactness}), simple non-quadratic regularizations
can be implemented by means of $\ell_p$ norms:
\begin{equation}
  \label{eq:Lp-norm}
  \Fprior(\Param) = \sigma \, \sum\nolimits_n q_n \, \Abs{\Param[n]}^{p} \, ,
\end{equation}
where $\sigma = \pm 1$ and $\V q \in \Reals^N_+$ are given nonnegative
weights.  For this regularization to be strictly convex, the weights must
all be strictly positive, $p > 0$ and $p \not= 1$ (taking $p =0$ or $1$ is
discussed in Section \emph{Sparsity Promoting Priors}) with $\sigma = +1$
if $p > 1$ and $\sigma = -1$ if $p < 1$. Under these conditions and
provided $\totalflux \not= 0$, the default solution is unique and applying
formula~(\ref{eq:separable-default}) it can be found to be:
\begin{equation}
  \Param[n]^\Tag{def} = \frac{
    \totalflux \, \Sign(c_n) \, \Abs{c_n/q_n}^{\frac{1}{p-1}}
  }{
    \sum_{n'} \Abs[\big]{c_{n'}^p/q_{n'}}^{\frac{1}{p-1}}
  } \quad (\forall n)\, .
\end{equation}
Hence $\Param^\Tag{def} \propto (\V c/\V q)^{1/(p-1)}$ where the division
and the exponentiation are done component-wise,  which shows that the shape
of the default image is driven by the weights $\V q$.  With $p=2$, the
default image in Eq.~(\ref{eq:default-quadratic-compactness}) is retrieved.
Taking $p = 1/2$ and $\sigma = -1$ to ensure strict convexity corresponds
to the \emph{square-root entropy}
\cite{Narayan_Nityananda-1986-maximum_entropy_review} given in
Eq.~(\ref{eq:mem-sqrt}).

A notable feature of the regularization by the weighted $\ell_p$ norm in
Eq.~(\ref{eq:Lp-norm}) with $p \in (0,1)$ (and $\sigma = -1$) is that it
acts as a strong barrier preventing the parameters from approaching zero.
This trick is a simple means to enforce positivity.

\subsection{Maximum entropy}
\label{sec:max-ent}

From a strict Bayesian point of view, the logarithm of the prior
probability is called the \emph{entropy} and the solution of the image
restoration problem in Eq.~(\ref{eq:constrained-problem}) is then the image
which maximizes the entropy while being compatible with the data.  However,
the name \emph{Maximum Entropy Methods} (or MEM for short) is usually
restricted to specific forms of the regularization penalty called the
\emph{negentropy} because it is the opposite of the entropy and originally
derived from information theory \cite{Jaynes-1968-prior_probabilities}.
Frieden \cite{Frieden-1972-maximum_entropy} was the first to apply MEM for
image restoration with the following regularization:
\begin{equation}
  \label{eq:mem-normal}
  \Fprior(\Param) = \sum\nolimits_n \Param[n]\,
  \log\Paren[\big]{\Param[n]/r_n} \, ,
\end{equation}
which assumes that strict normalization holds and where $\V r \in \Omega$
is a given \vocab{prior image}.  Without the normalization constraint, the
MEM regularization becomes \cite{Skilling_Bryan-1984-maximum_entropy}:
\begin{equation}
  \label{eq:mem-common}
  \Fprior(\Param) = \sum\nolimits_n \Brack[\big]{
    r_n - \Param[n]
    + \Param[n]\,\log\Paren[\big]{\Param[n]/r_n}
   } \, .
\end{equation}
Maximum entropy methods for image reconstruction from sparse visibilities
have been reviewed some time ago
\cite{Narayan_Nityananda-1986-maximum_entropy_review} and alternative
expressions for the negentropy are:
\begin{equation}
  \label{eq:mem-log}
  \Fprior(\Param) = -\sum\nolimits_n r_n\,\log(\Param[n]) \, ,
\end{equation}
or
\begin{equation}
  \label{eq:mem-sqrt}
  \Fprior(\Param) = -\sum\nolimits_n \sqrt{r_n\,\Param[n]} \, ,
\end{equation}
which have the property of imposing strict positivity of
the image (provided the prior image is also strictly positive).  In view of
the results shown by Fig.~\ref{fig:positivity?} and obtained with a simple
separable quadratic regularization, it is clear that strict positivity is
beneficial for interpolating the voids in the frequency coverage.

All these MEM regularizations are separable, but a non-separable variant
has been proposed by Horne \cite{Horne-1985} who introduced a
\emph{floating prior} defined by $\V r = \M R \cdot \Param$ with $\M R$ a
linear operator such that $\V r$ is a smoothed version of the image
$\Param$ or perhaps a version of it with some symmetries imposed.

The separable MEM regularizations are all strictly convex and, like the
$\ell_p$ norm with $p < 1$, impose that the sought image be strictly
positive everywhere ($\Param > 0$).  Using the results of Appendix
\emph{Separable Regularizations}, it can be shown that the prior image is
also the default solution: $\Param^\Tag{def} = \V r$.

\begin{figure*}
  \centering
  \includegraphics[width=170mm]{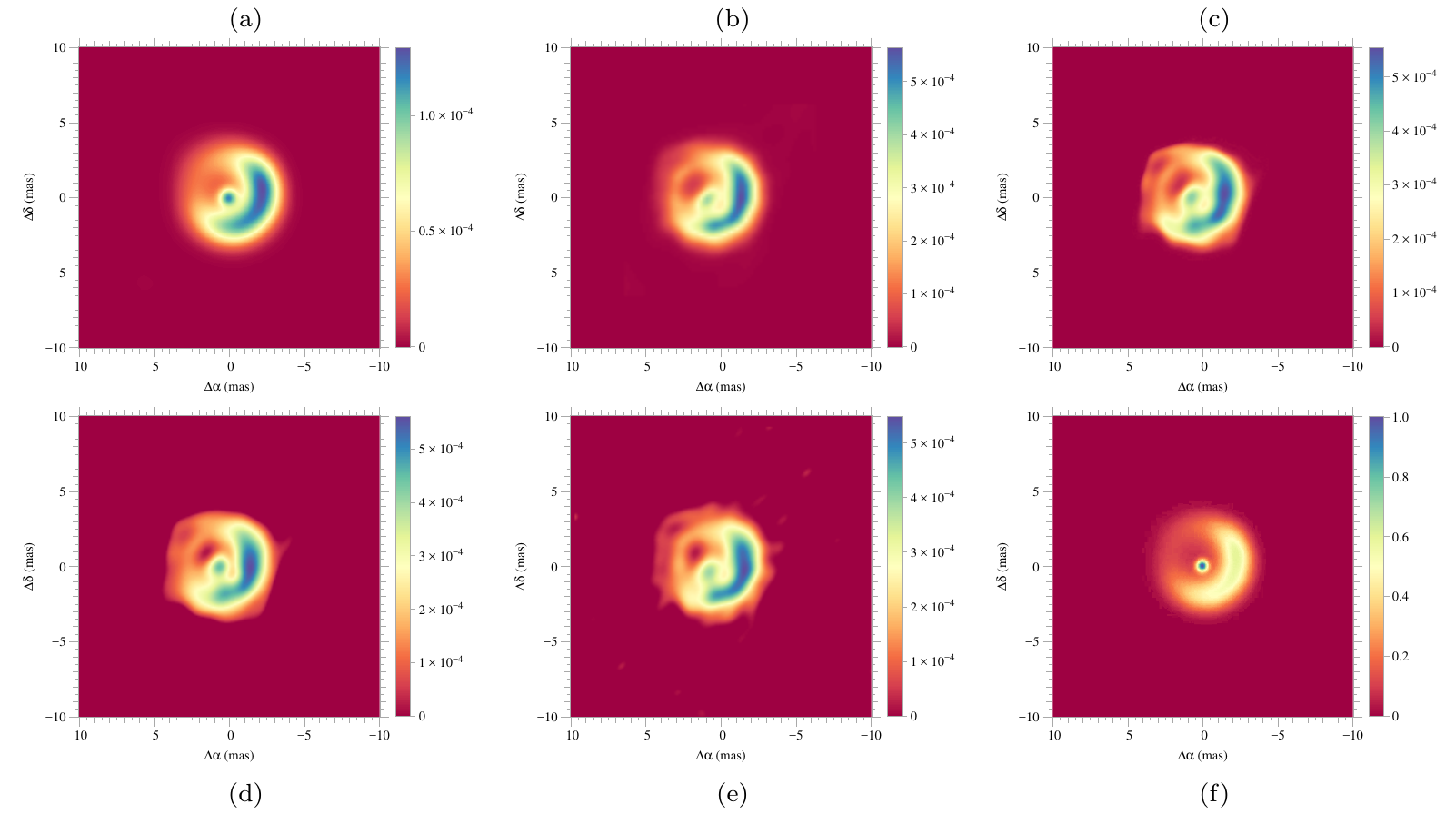}

  \caption{Reconstructions of LkH$\alpha$-101 with various regularizations.
  Panels: (a) original object smoothed at the resolution of the
  interferometer ($\mathrm{FWHM}\sim0.4\,\mathrm{mas}$); (b) \BSMEM
  reconstruction; (c) \MIRA reconstruction with MEM regularization as in
  Eq.~(\protect\ref{eq:mem-common}); (d) \MIRA reconstruction with a
  compactness quadratic regularization given by
  Eq.~(\protect\ref{eq:quadratic-compactness}); (e) \MIRA reconstruction
  with edge-preserving regularization implemented by
  Eq.~(\protect\ref{eq:hyperbolic-loss}) in
  Eq.~(\protect\ref{eq:smoothness}); (f) \SQUEEZE reconstruction with the
  $\ell_0$ norm of the \emph{à trous} wavelet coefficients.}

  \label{fig:various-regularizations}

\end{figure*}

\subsection{Sparsity promoting priors}
\label{sec:sparsity}

Seeking the most \emph{simple image} can be seen as seeking the image which
is explained by the fewest number of parameters.  This idea has led to the
\CLEAN algorithm \cite{Hogbom-1974-CLEAN} which attempts to fit the
interferometric data with the fewest point-like sources, possibly with an
extended smooth component which is added after recovering the point
sources, and perhaps under a support constraint.   Using the formalism of
Section \emph{Improving the Reconstruction}, the \CLEAN approach to recover
point-like sources could be approximated by:
\begin{equation}
  \label{eq:clean-method}
  \min_{\Param \in \Omega} \Norm{\V x}_0
  \quad\text{s.t.}\quad
  \Fdata(\Param) \le \eta \, ,
\end{equation}
where the pseudo $\ell_0$ norm $\Norm{\V x}_0$ simply equals the number of
non-zero pixels in the image $\V x$.  Minimizing the $\ell_0$ norm favors
\vocab{sparse solutions} where many parameters are exactly equal to zero.
Finding the global optimum of Eq.~(\ref{eq:clean-method}) is however
exceptionally difficult in terms of computational effort.  For instance, to
find a solution with $P$ non-zero parameters out of $N$ requires trying
each possible choice, that is:
\begin{displaymath}
  \binom{N}{P} = \frac{N!}{P!\,(N - P)!}
\end{displaymath}
possibilities. For a $N = 32\times32$ pixel image with $P = 10$ non-zero
pixels yields more than $3\times10^{23}$ possibilities!  Fortunately, many
works, both theoretical and practical, have shown that replacing the
$\ell_0$ norm by the $\ell_1$ one is almost as effective at selecting a
sparse solution which is, in many cases, a good approximation to the
sparsest solution compatible with the data \cite{Donoho-2006-ell1_approx}.
The enormous advantage of using the $\ell_1$ norm is that it leads to
convex optimization problems and, even though they are non-smooth, there
are many algorithms able to solve them exactly.  These properties of the
$\ell_1$ norm are the cornerstone of the development of so-called
\vocab{compressed sensing} methods
\cite{Candes_Wakin-2008-compressive_sampling}.


The sparsity principle can be extended to favor other structures besides
point-like sources.  To that end, it is assumed that the pixel values of
the image are given by $\V x = \M B \cdot \V z$ where $\V z$ are some other
parameters and the columns of $\M B$ form a \emph{dictionary} of structures
such that a few of them ought to describe the object of interest. Then the
problem to solve is:
\begin{equation}
  \label{eq:sparse-method}
  \min_{\V z} \Norm{\V z}_1
  \quad\text{s.t.}\quad
  \Fdata(\M B \cdot \V z) \le \eta
  \text{ and } \M B \cdot \V z \in \Omega \, ,
\end{equation}
which can be seen as an instance of the \noun{Lasso} algorithm
\cite{Tibshirani-1996-LASSO}.  The dictionary $\M B$ can be built from a
basis of wavelet functions \cite{Starck_et_al-1994-aperture_synthesis,
Baron-2016-image_reconstruction_overview} which has proven effective for
multi-scale structures.  Depending on the type of object, the result
obtained while imposing sparsity of the wavelet coefficients can be
impressive.  Figure~\ref{fig:various-regularizations} shows such an
example: the image reconstructed by \SQUEEZE is almost identical to the
true object shown in Fig.~\ref{fig:dirty-map}, whereas other methods fail
to reproduce the fine-scale structures accurately.

Note that if $\V x \in \Omega$ then $\V x \ge 0$ and $\Norm{\V x}_1 =
\sum_n \Abs{x_n} = \sum_n x_n = \totalflux$ which is a constant, thus the
least $\Norm{\V x}_1$ is not a useful regularization in this context.
\WISARD implements a so-called \emph{white $\ell_1$--$\ell_2$
regularization} which can be seen as a variant of the $\ell_p$-norm
regularizations.



\begin{figure*}
  \centering
  \includegraphics[width=170mm]{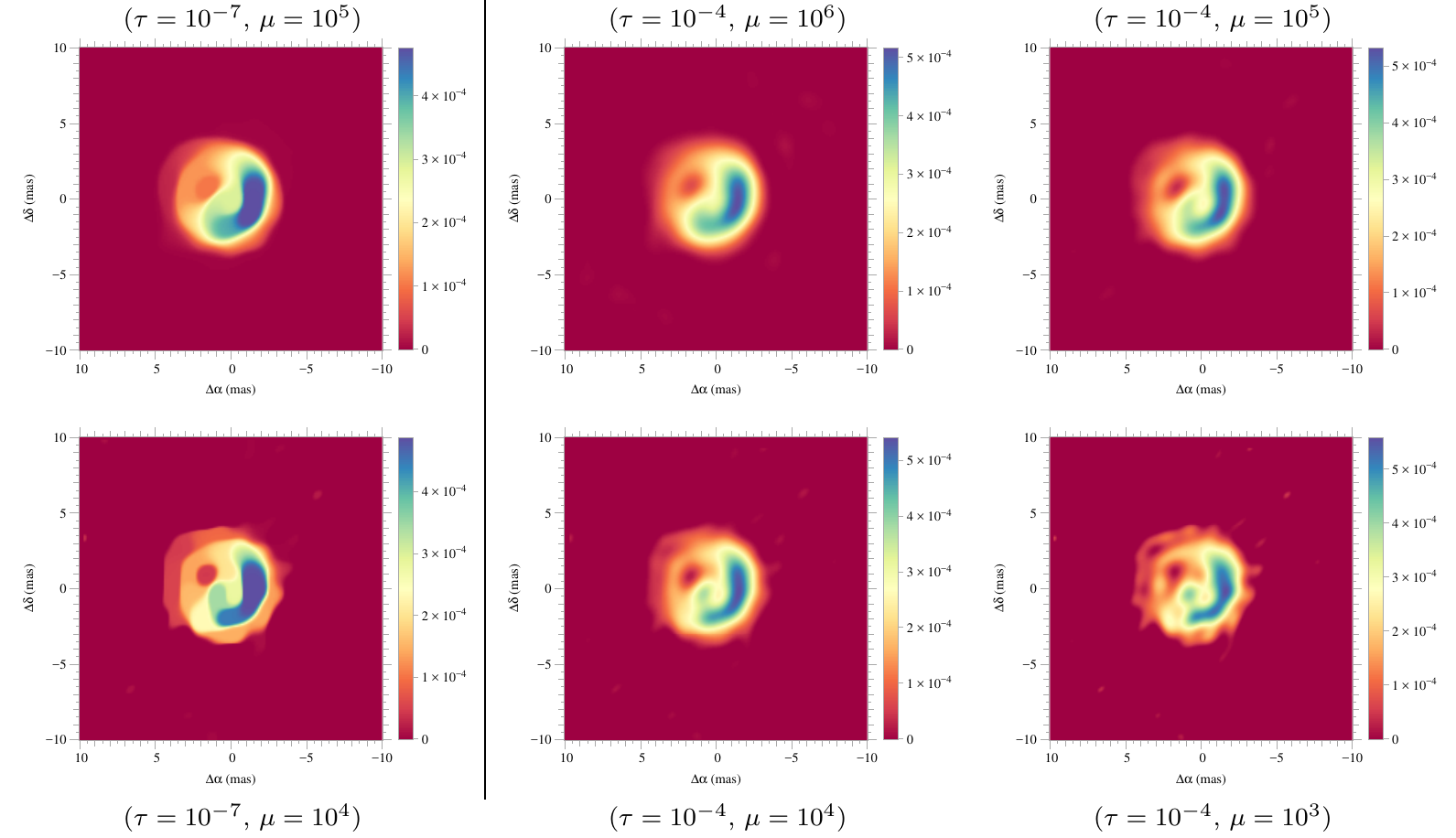}

  \caption{Reconstructions of LkH$\alpha$-101 with various regularization
  parameters. All images were obtained from the \emph{2004 Interferometric
  Beauty Contest} simulated data by \MIRA algorithm with the same
  edge-preserving regularization as in
  Fig.~\ref{fig:various-regularizations} but for different values of the
  hyper-parameters $\mu$ and $\tau$.  The images in the left panel have
  been obtained with a very small value of the threshold $\tau$ to mimic
  the behavior of TV regularization.}

  \label{fig:regularization-levels}
\end{figure*}

\subsection{Choosing the regularization}

We have seen that there are a variety of possible regularization terms
$\Fprior(\Param)$ that can be used. A comparison of some of the more
popular ones is presented in Fig.~\ref{fig:various-regularizations}.

In most practical cases, it is necessary to enforce positivity of the
reconstructed image, and to choose the weight of the regularization in
order to strike the correct balance between enforcing simplicity and
over-fitting the data. The weight can be treated as a nuisance parameter in
the Bayesian formalism, as is done by \BSMEM which tunes $\mu$ so that
$\Fdata(\Param)$ takes its expected value, corresponding to unit reduced
$\chi^2$. However, the optimal $\mu$ value can often be determined with
sufficient accuracy by visual inspection of the reconstructed images.
Fig.~\ref{fig:regularization-levels} clearly shows the effects of
over-regularization (which over-simplifies the image) and
under-regularization (which yields more artifacts).

A more sophisticated alternative to visual inspection for selecting the
optimum value for $\mu$ is the so-called ``L-curve'' method. This involves
computing the solution for, say 10 values of $\mu$, and plotting
$\Fprior(\Param)$ against $\Fdata(\Param)$. This relation should exhibit an
``L-shaped'' curve \cite{Hansen_1992}. For regions along the curve where
$\Fdata$ changes rapidly compared with $\Fprior$, the reconstructed image
is over-regularized. However, if the opposite is observed, the solution is
under-regularized. The elbow of the L-curve gives the optimum value of
$\mu$.

On the other hand, the choice of regularization function itself is a more
subjective issue. If little is known about the object to be reconstructed,
we recommend using several different regularizations in order to compare
their effects on the solution, and hence ensure that scientific conclusions
drawn from the images are robust.

In certain situations, prior knowledge about the object can be encoded in
the choice of regularization. For example, astrophysical considerations
tell us that a stellar disc (unless very cool) will have a sharp edge,
hence there is an objective justification for preferring regularizations --
such as edge preserving smoothness
\cite{Charbonnier_et_al-1997-edge_preserving} -- that favor sharp edges
over those that do not (\eg MEM).


\section{Image Reconstruction Algorithms}
\label{sec:algorithms}

We are now equipped to describe the image reconstruction methods that have
been successful when applied to realistic optical interferometric data and
which are sufficiently mature to be used with real data.  In addition to
coping with sparse Fourier data, these methods were specifically designed
to tackle the non-linear direct model of the data, to account for the
particular noise statistics  \cite{Meimon_et_al-2005-convex_approximation}
and to handle the OI-FITS data format \cite{Pauls_et_al-2005-oifits}. In
spite of this bias toward optical wavelengths, a method which can deal with
measured complex visibilities would be perfectly suitable to process
radio-interferometry data.

All the methods presented can be understood as an instance of an algorithm
to find a solution to one of the equivalent optimization problems
introduced in Section \emph{Improving the Reconstruction}. These methods
differ in the type of data which they take into account, in the expression
of $\Fdata(\Param)$ which measures the discrepancy between the data and the
model visibilities and in the assumed priors implemented via the
regularization function $\Fprior(\Param)$.  The number of possibilities
explains why there are so many different algorithms.  The many optimization
strategies exploited to solve the inverse problem also
contribute to the diversity of the algorithms. As not all strategies are
equivalent, before describing the image reconstruction algorithms, we give
some guidelines for understanding the benefits or the shortcomings of the
different optimization methods. A summary of the methods is given in
Table~\ref{tab:algorithms}.

\subsection{Optimization strategies}
\label{sec:optimization}

We have shown that image reconstruction amounts to solving a constrained
optimization problem like:
\begin{equation}
  \min_{\Param \in \Omega} \Brace[\big]{
     f(\Param) = \Fdata(\Param) + \mu\,\Fprior(\Param)
  } \, .
\end{equation}
Typically four kinds of strategies are used: gradient, augmented
Lagrangian, greedy and stochastic methods. Owing to the complexity of the
problem no closed-form solution exists and all these methods are iterative.

\vocab{Gradient-based optimization algorithms} update the variables in a
way that is inspired by Newton's method which requires that the objective
function $f(\Param)$ be at least twice continuously differentiable, hence
smooth enough. As there are too many variables to manipulate the whole
Hessian of the objective function, nonlinear conjugate gradients or limited
memory variants of variable metric methods have to be used
\cite{Nocedal_Wright-2006-numerical_optimization}. Such methods require the
computation of the objective function $f(\Param)$ and its gradient
$\nabla\!f(\Param)$. Gradient-based methods can be modified to account for
simple bound constraints (to implement the positivity) by means of active
sets. These methods yield a reduction of the objective function at each
iteration and their convergence is usually fast.  When applied to a
multi-modal objective function, they are however only able to find a local
optimum which is better than the initial solution.  In the context of
optical interferometry imaging, using non-linear observables such as the
power spectrum or the closure phase yields a multi-modal objective
function. A local optimum found by a gradient-based optimization method may
however provide an acceptable solution.  When only the power spectrum is
available, image restoration implies phase retrieval.  Projection methods
such as the Gerchberg-Saxton algorithm \cite{GerchbergSaxton1972} have been
used to solve phase retrieval problems but not in the context of stellar
interferometry where there is the added issue of the sparse frequency
coverage. Moreover, it has been shown that the Gerchberg-Saxton algorithm
is outperformed by gradient-based methods
\cite{Fienup-1982-phase_retrieval}.

When $f(\Param)$ is convex but not smooth, the optimization can be carried
out by an \vocab{augmented Lagrangian method} like the Alternating
Direction Method of Multipliers (ADMM)
\cite{Boyd_et_al-2010-method_of_multipliers} which exploits a variable
splitting strategy to separate the complex optimization problem into
sub-problems which are easier to solve, sometimes even exactly. ADMM also
provides a flexible way to impose the strict constraints implemented by
$\Omega$. ADMM has been successfully used in a number of algorithms for
interferometric imaging \cite{Giovannelli_Coulais-2005-pos_mix,
Thiebaut_et_al-2013-multispectral_sparsity, Schutz_et_al-2014-Painter,
Soulez_Thiebaut-2013_imaging_framework_OHP} but it requires as many control
parameters as there are splittings and constraints. These must be treated
as additional hyper-parameters needed to tune the algorithm.

The well known \CLEAN method \cite{Hogbom-1974-CLEAN,
Thompson_et_al-2017-radio_astronomy} is an example of \vocab{greedy
algorithm}.  Such algorithms, also called matching pursuit methods
\cite{Mallat_Zhang-1993-matching_pursuit}, are intended to solve a data
fitting problem under a sparsity constraint (see Section \emph{Sparsity
Promoting Priors}) expressed by the $\ell_0$ norm of the variables as in
Eq.~(\ref{eq:sparse-method}).  Greedy methods proceed by adding (or
removing) a single variable (\ie, a pixel of $\Param$ or a coefficient of
$\V z$ if a dictionary is used as in Eq.~(\ref{eq:sparse-method})) to the
current solution in order to improve the agreement with the data. This
strategy may be efficient when a very small number of non-zero variables
are sufficient to fit the data but they are nevertheless only able to find
a local solution which improves on the starting solution.  Note that the
efficiency of a greedy method strongly depends on the criterion for
selecting which variable to change.

The objective of \vocab{stochastic methods} is to find a close
approximation of the global minimum of the objective function $f(\Param)$.
Their name comes from the fact that they perform a random walk to explore
the objective function. During this exploration, a new candidate
$\underline{\Param}^{[t+1]}$ is generated by a random perturbation of the
current estimate $\Param^{[t]}$ and the new candidate is either kept or
rejected depending on how much it improves the solution compared to
$\Param^{[t]}$.  If there is an improvement, that is if
$f\Paren[\big]{\underline{\Param}^{[t+1]}} \le f\Paren[\big]{\Param^{[t]}}$
the new candidate is kept; otherwise the algorithm decides randomly whether
to keep $\underline{\Param}^{[t+1]}$ with a probability decreasing with the
value of $f\Paren[\big]{\underline{\Param}^{[t+1]}} -
f\Paren[\big]{\Param^{[t]}} \ge 0$.  If the new candidate is kept, then the
next estimate is $\Param^{[t+1]} = \underline{\Param}^{[t+1]}$; otherwise,
it is rejected and $\Param^{[t+1]} = \Param^{[t]}$.  By repeating this
procedure, a Monte Carlo Markov Chain (MCMC) is built
\cite{Geman_Geman-1984-Bayesian_restoration, MacKay-1998-MCMC}. In the case
of \emph{simulated annealing}, the probability of keeping a bad estimate is
slowly reduced so that the Markov chain cools down to the global optimum of
the objective function. Taking $f(\Param) =
-\log\PDF(\Param\given\Data,\Hyper)$ and without the cooling, MCMC is a
means to sample the posterior probability $\PDF(\Param\given\Data,\Hyper)$.
Then straightforward sample statistics can be applied to the elements of
the Markov chain to get the mode (that is the best solution found), the
posterior mean or other moments. Stochastic methods sound very appealing in
the context of optical interferometry because $f(\Param)$ is necessarily
multi-modal with non-linear observables. Another attractive feature is the
possibility to obtain not only an image but also error bars on the pixel
values.  Finally, as only the value of $f(\Param)$ is required,
non-differentiable and non-convex objective functions such as those based
on the $\ell_0$-norm may be considered. However, compared to local
optimization methods, stochastic methods take much more time to converge to
a solution and a fair amount of experience is necessary to chose the
parameters of a stochastic method correctly.

\begin{table*}
  \caption{\label{tab:algorithms} Summary of the algorithms for image
  reconstruction from optical interferometric data described in Section
  \emph{Image Reconstruction Algorithms}.}
  \centering
    \rowcolors{3}{tablerowcolor}{}
    \begin{tabular}{lP{8.1em}P{9.3em}ll}
      \TopRule
      \rowcolor{tableheadcolor}
      \bf Name & \bf Authors & \bf Optimization & \bf Regularization & \bf Multi-spectral \\
      \BSMEM & Baron, Buscher, Young & Trust region gradient & MEM-prior & No \\
      Building Block Method & Hofmann, Weigelt & Matching pursuit & Sparsity & No \\
      \IRBIS & Hofmann, Weigelt & ASA\_CG & Many & No \\
      \MACIM & Ireland, Monnier & Simulated annealing & MEM, darkness & No \\
      \MIRA & Thiébaut & VMLM-B & Many & No \\
      \WISARD & Meimon, Mugnier, Le~Besnerais & VMLM-B plus\\self-calibration & Many & No \\
      \SQUEEZE & Baron, Monnier, Kloppenborg & Parallel tempering & Many & Yes \\
      \SPARCO & Kluska \etal & VMLM-B & Many & Yes \\
      \PAINTER & Schutz \etal & ADMM & Many & Yes \\
      \MIRA-3D & Soulez \etal & ADMM & Many & Yes \\
      \BottomRule
  \end{tabular}
\end{table*}

\subsection{Algorithms for gray image reconstruction}

We first start by reviewing algorithms tailored for recovering a
monochromatic image. Assuming a \emph{gray object} (\ie one whose
appearance does not change with wavelength) or processing the spectral
channels independently, these methods can be used on multi-spectral data.

\newcommand*{\Item}[1]{\textit{#1)}}

\Item{1} \BSMEM algorithm \cite{Buscher-1994-BSMEM,
Baron_Young-2008-Marseille} makes use of a Maximum Entropy Method (Section
\emph{Maximum Entropy}) to regularize the problem of image restoration from
the measured bispectrum (hence its name).  The improved \BSMEM version
\cite{Baron_Young-2008-Marseille} uses the Gull and Skilling entropy, see
Eq.~(\ref{eq:mem-common}), and a likelihood term with respect to the
complex bispectrum which assumes independent Gaussian noise statistics for
the amplitude and phase of the measured bispectrum.  The optimization
engine is \noun{MemSys} which implements the strategy proposed by Skilling
\& Bryan \cite{Skilling_Bryan-1984-maximum_entropy} to solve the problem in
Eq.~(\ref{eq:constrained-problem}) and automatically finds the most likely
value for the associated Lagrange multiplier.  Because it makes no attempt
to convert the data into complex visibilities, a strength of \BSMEM is that
it can handle any type of data sparsity (such as missing closure phases).
Recently, \BSMEM has been updated to be usable from the unified Graphical
User Interface (GUI) developed in the framework of \noun{Opticon}
\cite{Young_et_al-2016-user_friendly}. \BSMEM contains proprietary
software, but is available for academic use subject to a no-fee license
agreement
(\url{http://www.mrao.cam.ac.uk/research/optical-interferometry/bsmem-software/}). The GUI is intended to be usable with any software which implements a specified common command line interface \cite{Thiebaut_et_al-2017-common_interface}, and is available at the \emph{Jean-Marie Mariotti Center} (JMMC, \url{http://www.jmmc.fr/oimaging}).

\Item{2} The Building Block Method
\cite{Hofmann_Weigelt-1993-building_blocks} is similar to the \CLEAN method
but designed for reconstructing images from bispectrum data obtained by
means of speckle or long baseline interferometry.  The method proceeds
iteratively to reduce a cost function $\Fdata^\BispectrumTag(\Param)$ equal
to that in Eq.~(\ref{eq:bispectrum-penalty}) with weights set to a constant
or to an expression motivated by Wiener filtering.  The minimization of the
penalty is achieved by a matching pursuit algorithm which imposes sparsity
of the solution.  The image is given by the building block model in
Eq.~(\ref{eq:grid-image-model}) and, at the $t$-th iteration, the new
image $I^{[t]}(\V\theta)$ is obtained by adding a new building block at
location $\V\theta^{[t]}$ with a weight $\alpha^{[t]}$ to the previous
image, so as to maintain the normalization:
\begin{equation}
  I^{[t]}(\V\theta)
   = (1 - \alpha^{[t]})\,I^{[t - 1]}(\V\theta)
  + \alpha^{[t]} \, h(\V\theta - \V\theta^{[t]}) \, .\notag
\end{equation}
The weight and location of the new building block is derived by minimizing
the criterion $\Fdata^\BispectrumTag$ with respect to these parameters.
Strict positivity and support constraints can be trivially enforced by
limiting the possible values for $\alpha^{[t]}$ and $\V\theta^{[t]}$. To
avoid super resolution artifacts, the final image is convolved with a
smoothing function with size set according to the spatial resolution of the
instrument.

\Item{3} \IRBIS algorithm \cite{Hofmann_et_al-2014-IRBIS} is an
evolution of the \emph{Building Block Method}.  It performs regularized
image reconstruction by fitting bispectrum data while imposing positivity
by means of the ASA\_CG \cite{Hager_Zhang-2006-active_set_algorithm} method
which is a non-linear conjugate gradient method with active set.  This
means that the result depends on the settings but also on the initial
image. \IRBIS implements various regularization methods: simple Tikhonov,
see Eq.~(\ref{eq:quadratic-prior}), Maximum Entropy, see
Eq.~(\ref{eq:mem-common}), and edge-preserving smoothness, see
Eq.~(\ref{eq:smoothness}) with the hyperbolic semi-norm given in
Eq.~(\ref{eq:hyperbolic-loss}).  The bispectrum data considered by \IRBIS
are synthesized from the power spectrum and closure phase data of an
OI-FITS file which provide the synthetic modulus and the phase of the
bispectrum respectively. The bispectra with one of the frequencies set to
zero are also included so that there is no loss of information.

\Item{4} \MACIM algorithm \cite{Ireland_et_al-2006-MACIM}, for MArkov Chain
IMager, aims at maximizing the posterior probability:
\begin{equation}
  \PDF(\Param \given \Data, \Hyper) \propto
  \exp\Bigl(-\frac{1}{2}\,\Fdata(\Param)
  - \frac{\mu}{2}\,\Fprior(\Param)\Bigr) \,.
\end{equation}
\MACIM implements MEM regularization and a specific \vocab{darkness prior}
via a regularizer which favors large regions of dark space in between
bright regions.  For this latter regularization, $\Fprior(\Param)$ is the
sum of all pixels with zero flux on either side of their boundaries. \MACIM
attempts to maximize $\Pr(\Param\vert\Data,\Hyper)$ by a simulated
annealing algorithm with Gibbs sampling
\cite{Geman_Geman-1984-Bayesian_restoration} which can in principle solve
the global optimization problem of maximizing
$\Pr(\Param\vert\Data,\Hyper)$, but convergence can be very slow,
especially for objects comprising multiple components.

\Item{5} \MIRA algorithm \cite{Thiebaut-2008-Marseille} defines the
sought image as the minimum of the penalty function in
Eq.~(\ref{eq:alternative-solution}).   Minimization is done by VMLM-B, a
limited variable memory quasi-Newton method (based on BFGS updates) with
bound constraints for the positivity \cite{Thiebaut-2002-optim_bdec}. \MIRA
is written in a modular way closely following the expression of the
likelihood in Eq.~(\ref{eq:fdata-heterogeneous}): any type of data can be
taken into account by providing a function that computes the corresponding
penalty and its gradient.  For the moment, \MIRA handles complex
visibility, power spectrum and closure-phase data via penalty terms given
by Eq.~(\ref{eq:fdata-local-approx}), Eq.~(\ref{eq:powerspectrum-penalty})
and Eq.~(\ref{eq:closure-penalty-von-Mises}). Like \BSMEM, \MIRA can cope
with any missing data, in particular, it can be used to restore an image
given only the power spectrum (\ie without any Fourier phase information)
with at least a $180^\circ$ orientation ambiguity.  An implementation of
\MIRA in \noun{Yorick}\cite{Munro-1995-Yorick} is freely
available (\url{https://github.com/emmt/mira}) which can be used
from the command line or from the \noun{Yorick} interpreter.

\Item{6} \WISARD algorithm \cite{Meimon_et_al-2005-weak_phase_imaging}
recovers an image from power spectrum and closure phase data.  It exploits
a self-calibration approach \cite{Readhead_Wilkinson-1978-VLBI,
Cornwell_Wilkinson-1981-self_calibration} to recover missing Fourier
phases.  Given a current estimate of the image and the closure phase data,
\WISARD first derives missing Fourier phase information in such a way as to
minimize the number of unknowns.  Then, the synthesized Fourier phases are
combined with the square root of the measured power spectrum to generate
pseudo complex visibility data which are fitted by the image restoration
step.  This step is performed by using the chosen regularization and a
quadratic penalty with respect to the pseudo complex visibility data, see
Eq.~(\ref{eq:fdata-local-approx}). This approach gives a unique solution
for the image restoration step, although overall the global problem remains
multi-modal. \WISARD has been implemented in IDL and can also be used with
GDL (\url{http://gnudatalanguage.sourceforge.net/}); it was originally
implemented at ONERA and is now maintained by the JMMC from where it is
freely available (\url{http://www.jmmc.fr/wisard_page.htm}).

\MIRA and \WISARD have been developed in parallel and share some common
features.  They use the same optimization engine
\cite{Thiebaut-2002-optim_bdec} and means to impose positivity and
normalization \cite{leBesnerais_et_al-2008-interferometry}.  However they
differ in the way missing data is taken into account: \WISARD
\emph{explicitly} solves for missing Fourier phase information; while \MIRA
\emph{implicitly} accounts for any lack of information through the direct
model of the data~\cite{leBesnerais_et_al-2008-interferometry}. Both
implement many different regularizers (negentropy, quadratic or
edge-preserving smoothness, compactness, total variation, \etc)

\subsection{Multi-spectral methods}

All current interferometers provide multi-spectral data and recovering a
multi-spectral image could be done in a naive way with one of the previous
algorithms by processing each spectral channel independently.  Jointly
processing the available data at all wavelengths is however much more
powerful for several reasons.  First it reduces the voids in the spatial
frequency coverage since the measured frequencies are wavelength dependent,
see Eq.~(\ref{eq:spatial-frequency}).  Second it offers the opportunity to
exploit the regularity of the specific brightness distribution of the
object along the spectral dimension which can significantly improve the
quality of the restored image
\cite{Bongard_et_al-2011-hyperspectral_deconvolution,
Thiebaut_et_al-2013-multispectral_sparsity}.  Using a monochromatic image
reconstruction algorithm and an image prior which is identical for the
different wavelengths, it is possible to retrieve a multi-spectral image of
the object \cite{leBouquin_et_al-2009-TLep} but exploiting the full
potential of multi-spectral imaging requires the development of specific
algorithms such as the ones described below.

\Item{7} \SQUEEZE algorithm \cite{Baron_et_al-2010-SQUEEZE} was
developed recently by Fabien Baron and John Monnier at the University of
Michigan, with the collaboration of Brian Kloppenborg from the University
of Denver. \SQUEEZE has a very comprehensive set of features, including
both Markov Chain Monte Carlo \cite{Geman_Geman-1984-Bayesian_restoration,
MacKay-1998-MCMC} (as in \MACIM) and gradient-based optimization engines,
and the ability to combine geometric model-fitting with model-independent
imaging.  Thanks to its powerful optimization strategies, \SQUEEZE can deal
with the most demanding regularizations such as those based on a truly
$\ell_0$ norm.  As a result, \SQUEEZE has a vast choice of priors: sparsity
via the $\ell_0$ or $\ell_1$ norms either separable or applied on wavelet
coefficients (various wavelet transforms are available), Total Variation,
Maximum Entropy, darkness, \etc \SQUEEZE is one of the few algorithms
capable of multi-spectral imaging. Novel capabilities such as imaging on
spheroids are also being implemented. A C implementation of \SQUEEZE is
freely available (\url{https://github.com/fabienbaron/squeeze}) and is
usable from the command line. The superiority of some \SQUEEZE results (see
\eg Fig.~\ref{fig:various-regularizations}) over other methods suggests
that $\ell_0$ sparsity priors can be very appropriate even though they
require a global optimization strategy.

\Item{8} \SPARCO algorithm \cite{Kluska_et_al-2014-SPARCO} is a
semi-parametric approach that has been proposed by Jacques Kluska to deal
with multi-spectral data.  The fundamental idea is to assume that the
astronomical object comprises several components whose specific brightness
distributions are separable functions of the angular direction and the
wavelength, say $I_c(\V\theta,\lambda) = F_c(\lambda)\,I_c(\V\theta)$ for
the $c$-th component and for $I_c(\V\theta)$ normalized.  Then the model
visibility at wavelength $\lambda$ and projected baseline $\V b$ is:
\begin{equation}
  V(\V\nu,\lambda) = \frac{
    \sum_c F_c(\lambda)\,V_c(\V\nu)
  }{
    \sum_c F_c(\lambda)
  }
\end{equation}
with $\V\nu = \V b/\lambda$ the frequency and $V_c(\V\nu) =
\FT{I}_c(\V\nu)$ the Fourier transform of $I_c(\V\theta)$.  It is further
assumed that the spectral energy distributions (SED) $F_c(\lambda)$ follow
simple power laws (which is justified for young stellar objects in the near
infrared).  In the case where there are two components, an unresolved star
and some extended environment, the model simplifies to:
\begin{equation}
  V(\V\nu,\lambda) = \frac{
    1 + (\lambda/\lambda_0)^p\,\xi\,V_\Tag{env}(\V\nu)
  }{
    1 + (\lambda/\lambda_0)^p\,\xi
  }
\end{equation}
where $\xi = F_\Tag{env}(\lambda_0)/F_\Tag{star}(\lambda_0)$ is the ratio
of the SED of the environment and of the star at some reference wavelength
$\lambda_0$,  $p$ is the difference between the spectral indexes of these
two SED and $V_\Tag{env}(\V\nu)$ is the normalized visibility of the
environment.  This model adds just two parameters ($\xi$ and $p$) to a
monochromatic image reconstruction method (the sought image being
$I_\Tag{env}(\V\theta)$ the normalized brightness distribution of the
environment) and \SPARCO has been implemented in \MIRA and in \SQUEEZE.

\Item{9} \PAINTER algorithm \cite{Schutz_et_al-2014-Painter} aims to
recover the specific brightness distribution of the object, that is a 3D
multi-spectral image which is the solution of an optimization problem like
the one in Eq.~(\ref{eq:alternative-solution}) except that the
regularization is replaced by the following two terms:
\begin{equation}
  \label{eq:Painter-prior}
  \mu\,\Fprior(\Param) = \mu_\Tag{angl}\,f_\Tag{angl}(\M D_\Tag{angl}\cdot\Param)
  + \mu_\Tag{sptrl}\,f_\Tag{sptrl}(\M D_\Tag{sptrl}\cdot\Param) \, .
\end{equation}
$\M D_\Tag{angl}$ are $\M D_\Tag{sptrl}$ are finite difference operators
along the angular and spectral dimensions respectively. The spatial
regularization is implemented by the function $f_\Tag{angl}$ and the
operator $\M D_\Tag{angl}$ while the spectral regularization is implemented
by the function $f_\Tag{sptrl}$ and the operator $\M D_\Tag{sptrl}$.  The
hyper-parameters $\mu_\Tag{angl} \ge 0$ and $\mu_\Tag{sptrl} \ge 0$ set the
relative importance of the regularity of the image along its spatial and
angular dimensions.  A variety of functions $f_\Tag{angl}$ and $\M
D_\Tag{angl}$ are available to impose quadratic, edge-preserving or total
variation smoothness.  \PAINTER solves the problem by an Alternating
Direction Method of Multipliers (ADMM)
\cite{Boyd_et_al-2010-method_of_multipliers}.  A
\noun{Julia} \cite{Bezanson_et_al-2017-Julia} implementation of
\PAINTER is freely
available (\url{https://github.com/andferrari/PAINTER.jl}).

\textit{10) \MIRA-3D algorithm}
\cite{Soulez_Thiebaut-2013_imaging_framework_OHP} by Ferréol Soulez is
designed for multi-spectral image reconstruction from optical
interferometric data.  The 3D multi-spectral image is the solution of the
problem given in Eq.~(\ref{eq:alternative-solution}) where a \vocab{joint
spatio-spectral} prior is imposed:
\begin{equation}
  \label{eq:joint-smoothness}
  \Fprior(\Param) = \sum_{n,\ell}
  \zeta\Paren*{\M D_{n,\ell}\cdot\Param
  } \, ,
\end{equation}
which is similar to the one in Eq.~(\ref{eq:smoothness}) except that
indexes $n$ and $\ell$ respectively run along the spatial and spectral
dimensions of the multi-spectral image $\Param$ and that $\M
D_{n,\ell}\cdot\Param$ is the 3D (spatio-spectral) gradient of $\Param$ at
the position of the spaxel $(\V\theta_n,\lambda_\ell)$.  Operator $\M
D_{n,\ell}$ expands as:
\begin{equation}
  \label{eq:spatio-spectral-gradient}
  \M D_{n,\ell} = \Matrix{r}{
    \M D^\Tag{angl}_{n,\ell} \\
    \alpha \, \M D^\Tag{sptrl}_{n,\ell}\\
  } \, ,
\end{equation}
where $\M D^\Tag{angl}_{n,\ell}$ and $\M D^\Tag{sptrl}_{n,\ell}$ are finite
difference operators along the the spatial and angular dimensions
respectively and the hyper-parameter $\alpha$ is tuned to adjust the
relative importance of the spatial and spectral gradients. Compared to the
separable regularization of \PAINTER, the advantage of the joint
spatio-spectral regularization proposed by \MIRA-3D is that, with $\zeta$
given by Eq.~(\ref{eq:hyperbolic-loss}) or Eq.~(\ref{eq:tv-loss}), it
favors the synchronization of strong spectral and spatial variations.  This
is consistent with an astronomical object made of distinctive components
with different spectra. Another original possibility offered by \MIRA-3D is
to reconstruct a temperature map of the object by assuming the SED of each
pixel is given by Planck's law for a black body
\cite{Soulez_et_al-2016-temperature_map}.

%
%
%
%

\section{Discussion}

We have presented the methods for image reconstruction from interferometric
data with particular attention to the specific issues pertaining at optical
wavelengths.  The existing algorithms can be understood in a common
framework where the inverse problem of image reconstruction amounts to
solving a constrained optimization problem.  The objective function to
minimize has two terms which reflect a balance between fitting the
data (via a likelihood term) and enforcing the priors (via a regularization
term). This balance may be adjustable via a so-called hyper-parameter. The
constraints are the normalization and the positivity of the image; the
latter behaves like a floating support which is very important to help
interpolate missing frequencies.

A range of existing algorithms for image reconstruction from optical
interferometry data are described. Most of them have been the challengers
for the successive \emph{Interferometric Imaging Beauty Contests}
\cite{Lawson_et_al-2004-image_beauty_contest,
Lawson_et_al-2006-image_beauty_contest, Cotton_et_al-2008-Marseille,
Malbet_et_al-2010-beauty_contest, Baron_et_al-2012-beauty_contest,
Monnier_et_al-2014-beauty_contest, Sanchez_et_al-2016-beauty_contest} which
have demonstrated that there is not a single best algorithm and that the
quality of the result also depends on the user's ability to select the
proper settings. Understanding the principles of image reconstruction is
therefore mandatory for choosing a particular algorithm and its parameters.
Part of the future success of optical interferometers  will depend on the
ability of astronomers to develop the skills needed to use the image
reconstruction tools correctly. In this regard, the present paper may be
useful.

Image reconstruction for interferometry is still a vivid domain of research
and development. New algorithms are being developed for multi-spectral
imaging. Efforts are being devoted to make existing algorithms more user
friendly \cite{Young_et_al-2016-user_friendly}. Global optimization by
means of stochastic methods seems a very promising approach
\cite{Baron-2016-image_reconstruction_overview} to avoid under-optimal
results and to implement non-convex sparsity constraints.

\section{Acknowledgments}

Éric Thiébaut wishes to thank Michel Tallon for fruitful discussions about
modeling the physical behavior of an interferometer. Acknowledgements are
also due to the two anonymous reviewers for their numerous comments and
suggestions which helped us to improve the paper and fix some weaknesses.

The research leading to these results has received funding from the
European Community's Seventh Framework Programme (FP7/2013-2016) under
grant agreement number 312430 (OPTICON).

\appendix
\section{Appendices}

\subsection{Penalty for angular data}
\label{sec:angular-penaly}

\newcommand*{\phase}{\phi}
\newcommand*{\phasedata}{\phase}
\newcommand*{\phasemodel}{\phase_{\Tag{mod}}}

We consider deriving a suitable form of the likelihood penalty for angular
data, like the closure phase given in Eq.~(\ref{eq:phase-closure}).
Consistency with the likelihood penalties for other kinds of data [see for
instance Eq.~(\ref{eq:fdata-Gaussian})] dictates that the phase penalty be
given by:
\begin{equation}
  f(\phasemodel) = c - 2\,\log[\PDF(\phasedata\given\phasemodel)] \, ,
  \label{eq:angular-penaly}
\end{equation}
where $\PDF(\phasedata\given\phasemodel)$ is the probability density
function of the measured phase $\phasedata$ conditioned to the knowledge of
the model phase $\phasemodel$ and $c$ is a constant which can be chosen so
that $f(\phasedata) = 0$.  There is however no consensus on
the form of the distribution $\PDF(\phasedata\given\phasemodel)$ and image
reconstruction algorithms may assume different expressions for the penalty
$f(\phasemodel)$ and hence different phase distribution.

As the considered phase is the argument of a complex number, it is only
defined modulo $2\,\pi$.  To account for this, Haniff \cite{Haniff1991}
proposed to define the phase penalty as:
\begin{equation}
  f_\Tag{Haniff}(\phasemodel)
  = \arc(\phasedata - \phasemodel)^2/\sigma^2_{\phase} \, ,
  \label{eq:Haniff-penalty}
\end{equation}
with $\sigma^2_{\phase}$ some estimation of the variance of the measured
phase and $\arc: \Reals \mapsto (-\pi,+\pi]$ a function which wraps its
argument in the interval $(-\pi,+\pi]$.  The above expression amounts to
approximating the distribution $\PDF(\phasedata\given\phasemodel)$ of the
wrapped phase by a truncated Gaussian.

It has been empirically found \cite{Thiebaut-2008-Marseille,
Schutz_et_al-2014-Painter} that a distance based on the phasors has a
better behavior with respect to numerical optimization than the penalty
proposed by Haniff.  This leads to the following penalty:
\begin{align}
  f_\Tag{phasor}(\phasemodel)
  &= \kappa \, \Abs[\big]{\mathe^{\mathi\,\phasemodel} - \mathe^{\mathi\,\phasedata}}^2
  \notag \\
  &= 2 \, \kappa \, \Brack[\big]{1 - \cos\Paren{\phasedata - \phasemodel}} \, ,
  \label{eq:von-Mises-penalty}
\end{align}
with $\kappa \ge 0$ a weight which has to be determined.  Imposing in
Eq.~(\ref{eq:angular-penaly}) that $f(\phasemodel) =
f_\Tag{phasor}(\phasemodel)$ (up to an additive constant) leads to:
\begin{displaymath}
  \PDF(\phasedata \given \phasemodel)
  \propto \mathe^{\kappa\,\cos\Paren{\phasedata - \phasemodel}} \, .
\end{displaymath}
The factor can be computed so as to normalize the distribution
$\PDF(\phasedata \given \phasemodel)$ on an interval of width $2\,\pi$ which
yields:
\begin{equation}
  \PDF(\phasedata \given \phasemodel) = \frac{
    \mathe^{\kappa\,\cos\Paren{\phasedata - \phasemodel}}
  }{
    2\,\pi\,I_0(\kappa)
  } \, ,
  \label{eq:von-Mises-pdf}
\end{equation}
with $I_n$ the modified Bessel function of the first kind and order $n$.
The above distribution is known as the von Mises distribution
\cite{Mardia-1975-directional_data_statistics}.  In order to derive the
value of the parameter $\kappa$, the following phase \emph{variance} can be
empirically estimated from the data and its expression can be computed
assuming von Mises distribution:
\begin{align}
  \sigma^2_{\phase}
  &= \Expect\Paren*{
    \Abs[\big]{\mathe^{\mathi\,\phasemodel} - \mathe^{\mathi\,\phasedata}}^2
   }
  \notag \\
  &= 2 \, \int_{-\pi}^{+\pi}
  \Brack[\big]{1 - \cos\Paren{\phasedata - \phasemodel}} \,
  \PDF(\phasedata \given \phasemodel) \, \mathd\phase
   \notag \\
   &= 2 - 2\,I_1\Paren{\kappa}/I_0\Paren{\kappa} \, .
   \label{eq:angular-variance}
\end{align}
This equation gives an implicit definition of the parameter $\kappa$ in the
penalty defined in Eq.~(\ref{eq:von-Mises-penalty}).   For a small angular
variance (or equivalently a large value of $\kappa$), the following
approximation holds:
\begin{equation}
  \kappa \approx 1/\sigma^2_{\phase} \, .
  \label{eq:approx-kappa}
\end{equation}
In addition and in the limits of small differences between the data and the
model phases, the two expressions in Eq.~(\ref{eq:Haniff-penalty}) and
(\ref{eq:von-Mises-penalty}) are equivalent and:
\begin{equation}
  f(\phasemodel)
  \approx (\phasedata - \phasemodel)^2/\sigma^2_{\phase} \, .
  \label{eq:approx-angular-penaly}
\end{equation}
Note that the
\emph{angular variance} $\sigma^2_{\phase}$ defined in
Eq.~(\ref{eq:angular-variance}) is exactly twice the value of the so-called
\emph{circular variance} of the wrapped phase
\cite{Mardia-1975-directional_data_statistics}.

\subsection{Separable regularizations}
\label{sec:separable-regularizations}

Separable regularizations take the form:
\begin{equation}
  \Fprior(\Param) = \sum_n f_n(\Param[n])
\end{equation}
where $\List{f_n: \Reals \mapsto \Reals}{n=1}{N}$ forms a family of
functions.  We assume that these functions are strictly convex.

With such a regularization function, the default solution is:
\begin{equation}
  \Param^\Tag{def} = \argmin_{\Param \in \Omega} \Fprior(\Param)
  = \argmin_{\Param \in \Omega} \sum_n f_n(\Param[n]) \, ,
\end{equation}
when the feasible set $\Omega$ is defined as in
Eq.~(\ref{eq:feasible-set}), the Lagrangian of this constrained problem is:
\begin{displaymath}
  \mathcal{L}(\Param;\alpha) = \Fprior(\Param) - \alpha\,\Inner{\V c}{\Param} \, ,
\end{displaymath}
with $\alpha \in \Reals$ the multiplier for the normalization constraint.
Taking into account the nonnegativity, the Karush-Kuhn-Tucker (KKT)
necessary conditions of optimality for a feasible solution are
\cite{Nocedal_Wright-2006-numerical_optimization}:
\begin{displaymath}
  \left\{\begin{array}{r@{}l}
    \forall n\text{, either }&
    \partial_{\Param[n]}\mathcal{L}(\Param;\alpha) = 0
    \text{ and }\Param_n \ge 0\,, \\
    \text{or }&\partial_{\Param[n]}\mathcal{L}(\Param;\alpha) > 0
    \text{ and }\Param_n = 0\,; \\
    \multicolumn{2}{l}{\Inner{\V c}{\Param} = \totalflux\,.}\\
  \end{array}\right.
\end{displaymath}
As $\partial_{\Param[n]}\mathcal{L}(\Param;\alpha) = f'_n(\Param[n]) -
\alpha\,c_n$ and if all $f_n'$ are continuous and bijective (which holds
for the $f_n$ are strictly convex and therefore their derivatives are
strictly monotonous) there is a unique default solution:
\begin{equation}
  x_n^+(\alpha) = \max\Brace[\big]{0,(f'_n)^{-1}(\alpha\,c_n)}
  \quad(\forall n)\,.
  \label{eq:separable-unscaled-default}
\end{equation}
It remains to find $\alpha$ so that $\Inner{\V c}{\V x^+(\alpha)} =
\totalflux$.  At least for the regularizations considered in this paper,
this turns out to be a simple task.  For instance in most cases, we found
that:
\begin{equation}
   \label{eq:homogeneous-separable-unscaled-default}
   x_n^+(\alpha') = \chi(\alpha'/\alpha) \, x_n^+(\alpha) \,,
\end{equation}
for some function $\chi: \Reals_+\mapsto\Reals_+$.  In other words, changing
$\alpha$ only change the normalization not the shape of $\V x^+(\alpha)$.  Then the default solution is simply:
\begin{equation}
  \DefSolution = \frac{\totalflux}{\Inner{\V c}{\V x^+(\alpha)}} \, \V x^+(\alpha) \,,
  \label{eq:separable-default}
\end{equation}
computed for any $\alpha > 0$.


\bibliography{OI_imaging_tutorial}
\end{document}